\documentstyle[aps,epsf,multicol]{revtex}

\begin{document}
\draft

\title{Echinocyte Shapes: Bending,
Stretching and Shear Determine Spicule Shape and Spacing}
\author{Ranjan Mukhopadhyay\footnote{Present Address:
Department  of Physics, University of Pennsylvania, Philadelphia,
Pennsylvania 19104}, Gerald Lim H.W., and  Michael Wortis \\
Department of Physics, Simon Fraser University\\
Burnaby, British Columbia, Canada V5A 1S6}

\maketitle

\begin{abstract}

We study the shapes of human red blood cells using continuum
mechanics.  In particular, we model the crenated,
echinocytic shapes and show how they may arise from a competition between
the bending energy of the plasma membrane and the stretching/shear
elastic energies of the membrane skeleton.  In contrast to earlier work, we
calculate spicule shapes exactly by solving the equations of continuum
mechanics subject to appropriate boundary conditions.
A simple scaling analysis of this competition reveals an elastic
length $\Lambda_{el}$ which sets the
length scale for the spicules and is, thus, related to the
number of spicules experimentally observed on the fully developed echinocyte.

\end{abstract}


\section{INTRODUCTION}

The normal resting shape of the human red blood cell (RBC or
erythrocyte) is a flattened biconcave disc (discocyte) about 8 $\mu$m
in diameter.
Treatment of erythrocytes in vitro with a variety of amphipathic
agents is known to transform this shape systematically and reversibly
into various other shapes such as
echinocytes (crenated shapes) and stomatocytes (cup-like shapes),
which are further subdivided into classes labelled I, II, III (Brecher
and Bessis, 1972; Bessis,
1973; Chailley et al. 1973; Mohandas and Feo, 1975).  In particular,
the echinocyte III is a more-or-less
spherical body covered evenly with 25-50 rounded protuberances,
which we shall call spicules.

These shape transformations---from the discocyte to the stomatocyte on one
side and from the discocyte to the echinocyte on the other---have
been studied experimentally for more than 50
years.  Understanding these shapes and shape transformations
is a classic problem in cell biology;
over the past three decades it has also attracted
the attention of physicists. What makes this problem so
intriguing is that the structure of the red cell is remarkably simple.
To a good
approximation it is simply a bag of fluid, a concentrated
solution of hemoglobin surrounded by a thin macroscopically
homogeneous membrane.  Thus, the diverse resting shapes which it exhibits can
be thought of as (locally) stable mechanical equilibrium
shapes of the membrane.  This membrane is composite (Steck, 1989;
Alberts et al., 1994).
On the outside is
the plasma membrane, a self-assembled fluid bilayer with a thickness
of about 4 nm, composed of a complex mix of
phospholipids, cholesterol, and dissolved proteins.
Lipids such as phosphatidylcholine, sphingomyelin, and glycophospholipids, 
which are neutral at physiological pH, 
are common in the outer leaflet, while 
phosphatidylserine and phosphatidylethanolamine 
predominate in the inner leaflet (Gennis, 1989; Alberts et al., 1994).
Due to the negative charge of phosphatidylserine, located
in the inner
leaflet, there is a significant difference
in charge between the two leaflets of the bilayer.
Inside the plasma membrane but linked to it by protein anchors is a
thin, two-dimensionally cross-linked protein cytoskeleton,
the membrane skeleton (Steck, 1989; Bennett, 1990).
The skeleton is an hexagonally linked net,
each unit of which is a filamentous spectrin tetramer
with an extended length of
about 200 nm.
The spectrin tetramer is negatively charged at physiological pH.
Thermal fluctuations reduce the projected distance
between the vertices of the net to about 76 nm, while the offset
between the plasma membrane and the spectrin network is 30--50 nm
(Byers and Branton, 1985; Liu et al., 1987).
Functionally, the
plasma membrane serves as an osmotic barrier,
passing water with relative
ease but controlling, via a system of pumps and channels, the flow of
ions and larger solute molecules. The membrane skeleton has the
role of supporting and toughening the plasma membrane, which would
otherwise disintegrate in circulatory shear flow.

While the RBC membrane is certainly heterogeneous at the length
scale of individual lipid molecules, lipid patches or 
the basic spectrin tetramer, on scales
longer than 100 nm it is reasonably homogeneous in its properties, and
it may make sense to treat it as a mechanical continuum.  In this
spirit and following early work by Canham (1970),
Helfrich (Helfrich, 1973; Deuling and Helfrich, 1976), and others,
we imagine replacing the (fluid-phase) plasma membrane by
an ideal uniform two-dimensional surface characterized by a bending
rigidity $\kappa_{b}$,
taken for the RBC to be (Waugh and Bauserman, 1995; Strey et al., 1995)
  $2.0 \times 10^{-19}$ J (roughly 50 $k_{B}T_{\rm room}$,
where $k_{B}$ is Boltzmann's constant),
a spontaneous curvature
$C_{0}$, which recognizes the outside-inside asymmetry of the leaflet
composition, and an area-difference-elasticity (ADE) term
(Helfrich, 1973; Evans, 1974; Svetina et al., 1985;
Bo\v{z}i\v{c} et al., 1992; Wiese et al., 1992;
Miao et al., 1994), which reflects the fact that a difference
in relaxed area, $\Delta A_{0}$, between
the outer and inner leaflets produces a ``bilayer couple''
(Sheetz and Singer, 1974)
tending to bend the membrane away from
the larger-area side (like a bimetallic strip).
We assign to the modulus $\bar \kappa$
associated with the ADE term a value
of (Waugh and Bauserman, 1995) $1.27 \times 10^{-19}$ J.
These terms taken
together constitute what is called the ADE model of the plasma membrane.
In addition, we represent the RBC membrane skeleton as a two-dimensional
elastic continuum characterized by a bulk stretching modulus
$K_{\alpha}$ and a shear modulus $\mu$ with
$\mu \approx 2.5 \times 10^{-6}$
J/m$^{2}$ and $K_{\alpha} \approx 3\mu$ (see discussion in Sec. V).
These two mechanical subsystems, constrained to the same mathematical
surface, will constitute our model of the
RBC shape at length scales above 100 nm (see Sec. II for full details).

The basic hypothesis of the mechanical approach is that the observed
RBC shapes must be shapes which minimize the energy subject to
appropriate constraints on volume and area,
i.e., that all the observed RBC shapes must emerge as local
energy minima of this model and that the observed shape transformations
must come about as a response of the minimum-energy shapes to changes
in the ``control parameters'' which characterize the model.  There are
relatively few such control parameters:  In addition to the known area and
volume of the RBC, we include in the list the three mechanical moduli
noted above, the spontaneous curvature $C_{0}$,
the relaxed area difference $\Delta A_{0}$ between the bilayer
leaflets,
and parameters describing the effective relaxed size and
shape of the membrane skeleton.
If all these control parameters were known or easily measurable, we could
predict the stable RBC shapes using continuum mechanics.
The fact that they are not restricts
us, as we shall see, to somewhat more generic predictions.

This program has already been carried out, albeit in somewhat
restricted form, with respect to the discocyte and stomatocyte shapes.
The shapes and shape transitions of laboratory-prepared fluid-phase
phospholipid vesicles (lacking any skeleton)
have been successfully described by the ADE model (Miao et al.,
1994; D\"{o}bereiner et al., 1997). In
particular, plausible discocytic and stomatocytic shapes do occur
as energy-minimizing shapes of the ADE model.
Furthermore, the bilayer-couple
mechanism (Sheetz and Singer, 1974) accounts qualitatively for many of the
chemically-induced discocyte-stomatocyte transitions observed in the
laboratory, in that stomatocytogenic agents tend to segregate to the inner leaf
of the bilayer, thus making $\Delta A_{0}$ negative and
favoring the invaginated
stomatocyte shape (see more below).
There has been discussion in the literature of the way
in which introducing the elastic energy of the
membrane skeleton modifies or improves the
detailed prediction for the shape of the normal resting discocyte
(Zarda et al., 1977; Evans and Skalak, 1980).
However, most treatments in the literature
tend to regard the cytoskeletal
elastic energy as providing only perturbative
modification of the basic shapes
which emerge from the pure-bending-energy model.  We shall argue that
this point of view fails completely in describing echinocyte shapes.

Until recently, it was a major difficulty
for a fully mechanical picture of the RBC shapes that
echinocytic shapes have not been found in the catalog of
minimum-energy shapes of the ADE model; however, several authors
have now pointed out what we believe is the correct resolution of
this problem (Waugh, 1996; Igli\v{c}, 1997; Igli\v{c} et al.,
1998a, b; Wortis, 1998): A pure-lipid vesicle with a
sufficiently positive spontaneous curvature
(or, equivalently, sufficiently large positive $\Delta A_{0}$)
adopts a vesiculated or
budded shape, in which one or more quasi-spherical buds appears on the
outside of the main body of the vesicle, attached to it via a narrow
neck or necks (see Fig. 1).  Such necks
are allowed as low-energy structures
for pure-lipid vesicles, because the bending energy depends on the sum
of the two principal curvatures (see Sec. II), so that large curvatures
of opposite sign (characteristic of necks) can still lead to a small mean
curvature and, thus, to low energy  (Fourcade et al., 1994).
But---and this is the crucial point---small
necks involve large elastic strains in the membrane skeleton and are,
thus, inhibited by elastic energy cost.  It is not surprising
that they are not seen for undamaged RBC's.  What our calculations
show is that the formation of echinocytes is driven by positive $C_{0}$
(equivalently, positive $\Delta A_{0}$) and that
the spiculated shapes seen in experiments can arise from a
competition between the bending energy of the plasma membrane, which
by itself would promote budding, and the
stretching and shear elastic energies of the membrane skeleton, which
prevent it.
To make this more
precise, we note that the ratio of the bending modulus to the shear
modulus (or the stretching modulus) defines an elastic length scale,
\begin{equation}
       \Lambda_{el}=\sqrt{{\kappa_{b}} \over \mu},
	 \label{Lambda}
\end{equation}
which yields $\Lambda_{el} \approx 0.28$ $\mu$m.  We shall see below
that this number sets the scale of spicule size and, thus, fixes the
spicule density of the fully developed echinocyte (echinocyte III).

The success of this approach strengthens
the hypothesis that RBC shapes can be
understood on the basis of continuum mechanics; however, in
themselves, such calculations
cannot answer the question, ``Why does a particular shape
occur under given experimental conditions?''
Such a question has two parts.
Why do the control parameters have the values they do?  And, how does
this set of control parameters lead to the observed shape?  The
second part is the proper domain of continuum mechanics.  The
first part is primarily biological or biochemical in content and
revolves around the mechanisms which control the
lipid composition of the leaflets of the plasma membrane, local
electrostatic effects, the composition and environment
of the spectrin network and its coupling to the plasma membrane, and
any other determinants of $C_{0}$, $\Delta
A_{0}$, the elastic moduli, and other control parameters.
Hopefully, this separation of the question is useful.

The plan of this paper is as follows.  In the remainder of this
section, we briefly review some history of echinocyte shapes and shape
calculations.  Section II presents the details of the
continuum-mechanics model.  Section III discusses the theory required
to solve the model numerically for echinocyte shapes, including a
full account of the scaling argument
suggested above.  Section IV presents
our results for the predicted echinocyte shapes and spicule density.
Section V contains discussion of our results and further speculations.

The discocyte-to-echinocyte transformation was precisely identified by
Ponder (1948, 1955) and has since been studied
extensively by many other authors, as reviewed, for example, by Bessis
(1973), Lange et al. (1982), and Steck (1989).
Although this transformation can be driven
in many ways, the final spiculated shapes produced are apparently the
same (Smith et al., 1982; Mohandas and Feo, 1975).
This 
quasi-universal behavior 
is evidence for the kind of mechanism suggested
above, in which quite diverse biochemical processes
may set (a few) important
control parameters to the same or similar values.
In the same spirit, we note that
RBC ghosts (formed by hemolysis) behave quite similarly to
intact red cells (Lange et al., 1982).

The simplest picture of this type---consistent with
some (but not necessarily all) experimental findings---is
the so-called bilayer-couple
hypothesis of Sheetz and Singer (1974),
which focuses attention on
the control parameter $\Delta A_{0}$.
Thus, adding exogenous phospholipids to the exterior
solution is observed to promote echinocyte formation
(Ferrell et al., 1985; Christiansson et al., 1985).
This is explained in the
bilayer-couple picture by
arguing that such addition
is expected to produce an area
increase in the outer leaflet only, since phospholipids do not readily
flip-flop from one leaflet to the other.  This increases the area
difference $\Delta A_{0}$ and is expected
to promote echinocyte formation.  Similarly, cholesterol
does equilibrate across the bilayer but apparently prefers the outer
leaflet, presumably because of the particular lipids there present.
Thus, adding cholesterol will tend to increase $\Delta A_{0}$ and
promote echinocytosis, while depleting cholesterol will tend to
decrease $\Delta A_{0}$ and to promote stomatocytosis (Lange and
Slayton, 1982).
More generally, anionic
amphipaths all tend to produce echinocytes, while cationic amphipaths
tend to produce stomatocytes (Deuticke, 1968; Weed and Chailley, 1973;
Mohandas and Feo, 1975; Smith et al., 1982).
The bilayer-couple explanation of these observations is that,
when incorporated into the
plasma membrane, the cationic compounds
segregate preferentially to the inner
leaflet and the anionic compounds, to the outer leaflet because of
the predominantly negative charge of the inner-leaflet lipids.
Outer-leaflet segregation increases $\Delta A_{0}$, thus promoting
echinocytosis; conversely, inner-leaflet segregation
promotes stomatocytosis.
Time-dependent effects have even been seen, where a molecule
initially intercalates into the outer leaflet, causing
echinocytosis, but then slowly migrates to the
inner leaflet, following the electrostatic gradient,
causing return to the discocyte and subsequent
stomatocytosis (Isomaa et al., 1987).

A similar hypothesis is made to explain the observed tendency of ghosts
to become echinocytes at high salt concentrations but stomatocytes at
low salt concentrations (Lange et al., 1982).  The argument
relies on electrostatics.   Cationic species
neutralize the negative lipid charges on the inner leaflet and salt
decreases the Debye length, more effectively screening such charges.
Both effects lead to a contraction of the inner leaflet and consequent
increase in $\Delta A_{0}$, promoting crenation.

Other observations can be linked to the bilayer-couple effect but
somewhat less directly.  Thus, it is found that hemolyzed echinocytes
produce discocytic ghosts (Lange et al., 1982) and that
electroporation suppresses shape changes (Schwarz et al., 1999a, b).
To explain
these effects, it is argued that hemolysis and electroporation both
involve perforation of the membrane and resulting in contact via the
pore surface between the lipids in the inner and outer leaflets.
This provides a
mechanism for lipid transport which may reduce
the magnitude of $\Delta A_{0}$ 
by
partial symmetrization of the lipid composition of the leaflets, thus
(equivalently) reducing 
$C_{0}$.
Similarly, it is known that ATP depletion drives discocytes to
crenate (Nakao et al., 1960, 1961, 1962; Backman, 1986).
The hypothesis here is that ATP is required in some way to
maintain the lipid asymmetry of the leaflets (related to $C_{0}$),
perhaps for the
operation of ATP-driven translocases (Steck, 1989).

Some effects are not readily explained by the
bilayer-couple mechanism. 
Weed and Chailley (1972), and Gedde et al. 
(1995, 1997a, 1997b, 1999) report that
RBC shape can be controlled experimentally by varying the external pH,
with high pH promoting
echinocyte formation and low pH, stomatocyte formation (the effect
of proximity to a glass surface in promoting echinocytosis
(Furchgott and Ponder, 1940) is
probably related to this effect).
The mechanism for this effect does not seem to be well established.
Some authors argue for a mechanism involving
membrane proteins.
Band 3 has been specifically implicated (Gimsa and Ried, 1995; Jay,
1996; Gimsa, 1998;
H\"{a}gerstrand et al., 2000).
For example, it is known that high pH
induces dissociation of ankyrin and Band 3, which plays a role in the
linkage of the membrane skeleton to the bilayer (Low et al., 1991).
Indeed, it was shown that in vitro
the membrane skeleton expands at
high pH and contracts at low pH (Elgsaeter et al., 1986; Stokke et al., 1986).
It has recently been
proposed that pH change may induce conformational transformation in band 3
protein, leading to a change in $\Delta A_{0}$.  Similarly, Wong
(1994, 1999) has proposed that the folding of the spectrin in the
cytoskeletal net is controlled by a conformational
change of the Band 3 protein.  Such mechanisms might
shift the emphasis of shape determination from the dominantly
lipid-related control parameters, $C_{0}$ and  $\Delta A_{0}$, to the
elastic parameters and the relaxed shape of the membrane skeleton.  We
shall comment briefly on these issues in Sec. V.

Numerous shape calculations have previously been done using variants
of the model we describe in Sec. II.  Early calculations (Fung and
Tong, 1968; Zarda et al., 1977;
Evans and Skalak, 1980) omitted the
ADE term ($\bar \kappa =0$), set $C_{0}=0$,
and concentrated entirely on the
shape of the resting discocyte and the modifications caused by
osmotic swelling.  They solved numerically the
full mechanical shape equations.
Landman (1984) treated a viscoelastic shell
surrounding a viscous droplet and modeled the formation of protrusions
as sudden local addition of shell material.
Waugh (1996) studied a full ADE
model with local shear elasticity (but treating the compressibility
modulus as infinite) and studied the instability of a flat membrane to
formation of a local ``bump'' (axisymmetric spicule)
for positive values of $C_{0}$ or $\Delta A_{0}$ or both. He assumed a
specific spicule-shape parametrization, focussed on a
single spicule, calculated its energy, and in this way was able
to discuss generically the conditions under which spicule formation
might become energetically favorable.  Most recently, Igli\v{c} and
collaborators (Igli\v{c}, 1997; Igli\v{c} et al., 1998a, b) 
have used the
same model as Waugh (1996) but parametrize the spicule more simply, using a
hemispherical cap on a cylindrical body and joining this to a spherical
surface via a base shaped like a toroidal section.
They correctly identify the emergence of echinocyte shapes as a
competition between the bending energy and the skeletal elasticity.
They model the
echinocyte as a collection of $n_{s}$ spicules attached to a spherical
body and then determine the radius of the sphere, the parameters
of the spicules, and the preferred number of spicules by minimizing
the mechanical energy subject to constraints on cell area and volume.
In this way, they are able to estimate the expected number of
spicules and their shape as a function of the relaxed area
difference $\Delta A_{0}$.

The present paper builds on the work of
Waugh and Igli\v{c}:  We extend the model to include the area modulus
$K_{\alpha }$ of the membrane skeleton.  We calculate spicule shape
for the first time from the Euler-Lagrange equations (these
calculations are closely related to the earlier work of Zarda et al. (1977)).
Although we continue with Igli\v{c} et al. (1998a, b) to
treat the
echinocyte as a sphere decorated with spicules, we deal seriously
(although still approximately) with the mechanical boundary conditions
which must be met where the spicule joins the sphere.

\section{THE MODEL}

The central assumption of our work is that the
red blood cell assumes a shape that minimizes its overall membrane
energy subject to the appropriate constraints. The RBC
membrane is a composite of the plasma membrane and the cytoskeletal
network; correspondingly, we take the membrane
energy to be a sum of two terms,
\begin{equation}
     F=F_{b}+F_{el},
     \label{totalenergy}
\end{equation}
the bending energy,
\begin{equation}
F_{b}={1\over2}\kappa_{b}\int dA(C_{1}+C_{2}-C_0)^{2}
    + {{\bar \kappa} \over 2} {\pi \over A D^{2}}(\Delta A
                    - \Delta A_{0})^{2},
\label{bendingenergy}
\end{equation}
associated with the bilayer and
an elastic energy of stretching and shear,
\begin{eqnarray}
    F_{el}= {1 \over 2}K_{\alpha} \int dA_{0} (\lambda_{1}\lambda_{2}
- 1)^{2} +{\mu \over 2}\int dA_{0} \left({{\lambda_{1}}\over{\lambda_{2}}}
              +{{\lambda_{2}} \over {\lambda_{1}}} -2 \right),
\label{elasticenergy}
\end{eqnarray}
associated with the skeleton.
In doing this, we treat the plasma membrane as incompressible.
An estimate of its stretching modulus is
$K_{\alpha}^{(\rm bilayer)} \sim 10^{-1}$ J/m$^{2}$, comparable to that
of a pure-phospholipid bilayer
and some four orders of magnitude larger
than that of the skeleton.  Correspondingly, we ignore any
bending rigidity associated with the
isolated membrane skeleton (the dimensional
estimate $\kappa_{b}^{(\rm skeleton)} \sim K_{\alpha}
\times$(thickness)$^{2}$
leaves it two orders of magnitude smaller than that of the bilayer).

Eq.~\ref{bendingenergy} for the plasma
membrane's contribution to the overall membrane energy
is the now-standard ADE Hamiltonian
(Svetina et al., 1985;
Bo\v{z}i\v{c} et al., 1992; Wiese et al., 1992;
Miao et al., 1994).  The
first term was originally proposed by Helfrich (Helfrich, 1973;
Deuling and Helfrich, 1976).
$C_1$ and $C_2$ are the two local principal curvatures, $C_0$
is the spontaneous curvature, and the integral is over
the membrane surface.
The two leaflets of a closed bilayer of fixed interleaflet
separation $D$ are required by geometry to differ in area by an
amount,
\begin{equation}
    \Delta A = D \int dA(C_{1}+C_{2}).
    \label{areadifference}
\end{equation}
In calculations, we shall take $D \approx 3 \rm nm,$ corresponding to
the distance between the neutral surfaces of the leaflets.
The second term in Eq.~\ref{bendingenergy}
  is the so-called
area-difference-elasticity energy and represents the cost in
stretching (or compressional) energy of the individual
leaflets necessary to force
the change from the
relaxed area difference $\Delta A_{0}$ so as to conform to this geometric
requirement.  This effect occurs because a strong hydrophobic barrier
prevents lipids in one leaflet from ``flip-flopping''
passively to the other
on the time scales of mechanical shape changes.
$A$ is the area of the plasma membrane.
The modulus ${\bar \kappa}$ associated with the ADE term is
generically  of the same scale as $\kappa_{b}$.
Note, finally, that we can substitute Eq.~\ref{areadifference}
to rewrite Eq.~\ref{bendingenergy},
\begin{eqnarray}
	F_{b}={1\over2}\kappa_{b} \left[
	        \int dA(C_{1}+C_{2})^{2}
	        +{{\pi \alpha} \over A}\left( \int
	        dA(C_{1}+C_{2})\right)^{2}
	      -2{\bar C}_{0}\int dA(C_{1}+C_{2}) \right] +{\rm const.},
	\label{bendingenergy1}
\end{eqnarray}
where
\begin{equation}
	{\bar C}_{0}=C_{0}+{{\pi \alpha \Delta A_{0}}\over {DA}},
	\label{C0bar}
\end{equation}
$\alpha ={\bar \kappa}/\kappa _{b}$, and the constant term is shape
independent.  This shows that, in determining
minimal shapes at given values of
the control parameters, $C_{0}$ and
$\Delta A_{0}$ do not enter independently
but only in the form of an effective
spontaneous curvature ${\bar C}_{0}$
or, equivalently, an effective relaxed area difference,
\begin{equation}
	\overline{\Delta A}_0 = \Delta A_0 + {DAC_0 \over {\pi\alpha}},
	\label{DeltaA0bar}
\end{equation}
which we shall often quote in the dimensionless, reduced form
\begin{equation}
	\overline{\Delta a}_0 = {\overline{\Delta A}_0 \over A}
	=\Delta a_0 + {DC_0 \over {\pi\alpha}}.
	\label{Deltaa0}
\end{equation}
Note for future reference that increasing ${\Delta a}_{0}$ by $0.01$
is equivalent to increasing $C_{0}$ by $6.7 \mu$m$^{-1}$.

Eq.~\ref{elasticenergy} measures the
elastic-energy cost of the spectrin network.
It depends both on the relaxed shape of the membrane
skeleton and on the way it is actually distributed over the membrane
surface.  (The notion of a relaxed shape is, of course, somewhat
nominal, since removing the skeleton from the plasma membrane--which
can certainly be done (Sheetz, 1979)
--would radically
change its local biochemical environment, in a way which would modify
its shape and elastic constants).  In this redistribution, each element
of the network
will be stretched or compressed.  The quantities $\lambda_{1}$
and $\lambda_{2}$ are the local principal extension ratios of each membrane
element (Evans and Skalak, 1980).
$K_{\alpha}$ and $\mu$ are the (two-dimensional) moduli for
stretch and shear, respectively, as introduced in Sec. I, and the
integrals are over the undeformed shape.
In writing Eq.~\ref{elasticenergy}, we are assuming, as appears to be
typical, that the membrane skeleton does not plastically deform
in the course of the experimental shape transformation being
described. Note that Eq. \ref{elasticenergy}
makes a particular choice of terms in the elasticity beyond
those quadratic in the weak-deformation parameters,
$\epsilon_{i}=\lambda_{i}-1$.  For the large elastic
strains which are present at narrow necks and
for small spicules, these nonlinearities
do play a role.  As far as we are aware,
RBC elasticity has not been measured well enough to produce a clear
preference for a particular form of the nonlinearities.
Various authors have used
Eq.~\ref{elasticenergy} for
the elastic energy in other contexts with apparent success;
however, alternative forms of the elasticity have also been proposed
for dealing with problems involving large local deformations 
(Evans and Skalak, 1980; Discher et al., 1994).
We shall make some additional comments in Sec. V.

One attractive feature of Eq. \ref{elasticenergy} is that
the effect of changing the size of the relaxed membrane skeleton
by a pure, uniform dilation is particularly simple.
Suppose that the skeleton is decreased in linear scale by a factor $b$.
This means that undeformed areas are reduced by a factor $1/b^{2}$,
while the extension ratios $\lambda_{1,2}$ are each increased by a factor
$b$.  Note that the integrand of the shear term
in Eq.~\ref{elasticenergy} is invariant under this change but the factor
$\lambda_{1}\lambda_{2}$ in the stretching integrand increases by $b^{2}$.
Only the quadratic part of the stretching term influences the
membrane shape because $\int dA_{0}\lambda_{1}\lambda_{2}$ is just
the membrane area A, which is fixed by the incompressibility of the
plasma membrane.  It follows that the effect of decreasing the
size of the skeleton is completely equivalent to keeping the size
fixed but, instead, changing the elastic moduli according to
\begin{eqnarray}
        K_{\alpha}^{'}&=&b^{2}K_{\alpha} \label{cytoscaling0} \\
        \mu^{'} &=& b^{-2}\mu,
	  \label{cytoscaling}
\end{eqnarray}
\noindent i.e., this ``prestress''
in the membrane skeleton is completely equivalent to no prestress, a harder
stretching modulus, and a softer shear modulus.  It is not known with
certainty what ``prestress'' exists in the typical RBC skeleton.
Computer simulation has suggested that the relaxed skeleton may be as
much as 10--20\% smaller in area than the plasma membrane (Boal, 1994).
On the other hand, the experiment by Svoboda et al. (1992) shows
that isolated skeletons are expanded.
In our calculations, 
we shall assume zero net prestress (relaxed skeletal area equal to RBC
area, i.e., $\gamma \equiv A/A_{0}=1$).
Any deviation from this would result in modified, effective
elastic constants according to Eqs.~\ref{cytoscaling0} and \ref{cytoscaling}.

In our calculations, we will identify shapes which are minima of
the (free) energy defined by
Eqs.~\ref{totalenergy}--\ref{elasticenergy} subject to constraints of
fixed membrane area $A$ (bilayer incompressibility)
and volume $V$ (since the
RBC volume is typically set by osmotic balance).
Constrained minimization is achieved variationally
by introducing the functional,
\begin{equation}
        \Phi(\Sigma,P)= F + \Sigma A - P V,
	  \label{Phi}
\end{equation}
where $\Sigma$ and $P$ are Lagrange multipliers used to enforce
the surface-area and volume constraints. $P$
is the pressure difference across the bilayer, while $\Sigma$
has the dimension of a surface tension. Making $\Phi$ stationary
with respect to variations of membrane shape and cytoskeletal
distribution
leads to a set of coupled Euler-Lagrange equations.
These equations can then be solved to give the shapes of mechanical
equilibrium.
In principle, such
shapes are expected to correspond to observed equilibrium shapes
at temperature $T=0$ only; however,
because the energy scale $\kappa_{b}$ is large on the scale of $k_{B}T$,
thermal fluctuations are generally negligible and play an
important role only for ``soft'' modes,
especially near instabilities (Wortis et al., 1997).
The Euler-Lagrange equations are in general nonlinear and can have
multiple solutions.  The lowest-energy solution for given $A$ and $V$
is automatically stable to small fluctuations; higher-energy
solutions must be tested for stability.  All local-minimum solutions
are candidates for stable observable shapes, except in exceptional cases
(near instability) where energy barriers become comparable to $k_{B}T$.

The stable shapes produced by this process depend on the control
parameters: the geometrical parameters, $A$ and $V$; the determinants
of curvature, $C_{0}$ and $\Delta A_{0}$, which only enter in the
combination (\ref{C0bar}) or, equivalently, (\ref{DeltaA0bar})-(\ref{Deltaa0});
the moduli, $\kappa$, $\bar
\kappa$, $K_{\alpha}$, and $\mu$; plus, finally, any parameters required to
characterize the relaxed shape of the membrane skeleton.  The area 
and volume of
a typical RBC we take as (Bessis, 1973) $A=140 \mu$m$^{2}$ and $V=90
\mu$m$^{3}$.  The moduli, given in Sec. I, are less well determined by
experiment (we shall have more to say on this point in Sec. V).  This
leaves as unknown control parameters the curvature variables
and the cytoskeletal parameters.

The Euler-Lagrange variational equations derived from (\ref{Phi}) are
not numerically tractable except in the case of axisymmetry, which
clearly does not apply to echinocytic shapes.  In this paper, we
aim to treat only fully developed echinocytic shapes (echinocyte
III).  For this case, individual spicules are identical and
axisymmetric in shape to a good
approximation and, in addition, the central body which they decorate
is approximately spherical with radius $R_{0}$.
The observed distribution of spicules is
rather regular, and we shall approximate the local spicule packing as a
triangular array (except for special
numbers of spicules, there must, of course, be some defects), which
will look increasingly like an hexagonal
close-packed structure, as the number of spicules, $n_s$, becomes large.

The base of each individual spicule, where it meets the sphere
tangentially, is a circular contour $L$ of radius $r_{L}$.
If $\theta_{L}$ is
the angle subtended by $L$ at the center of the sphere, then
\begin{equation}
	r_{L}=R_{0}\sin \theta_{L}.
	\label{geometry}
\end{equation}
Because of the
close-packed structure, the circles $L$ meet tangentially.
We explain in Sec. III how to
derive appropriate boundary conditions where the
spicules meet the sphere.
Solving the axisymmetric Euler-Lagrange equations then determines a
spicule shape, including an individual spicule
volume $V_{s}$ and area $A_{s}$.  In
terms of these variables, the overall area and volume of the RBC are
taken to be
\begin{equation}
	A \approx nA_{s} + 0.1\times 4 \pi R_{0}^{2}
	\label{area}
\end{equation}
and
\begin{equation}
	V \approx {4\pi \over 3}R_{0}^{3} + nV_{s},
	\label{volume}
\end{equation}
with the spicule number
\begin{equation}
	n_{s} \approx 3.6 \times \left( R_{0} \over r_{L} \right)^{2}.
	\label{spiculenumber}
\end{equation}
In writing these relations, we
have assumed that 10\% of the spherical surface is not
covered by the circular spicule bases (close packing on a flat surface
would give 9.3\%; curvature effects and packing defects both increase this
number).  The spicule volume $V_{s}$ is calculated with respect
to a plane through $L$, and Eq. \ref{volume} neglects curvature
corrections.  These approximations are good when the number of spicules is
large, as it will turn out to be for the echinocyte shapes.

\section{THEORY}
\subsection{Membrane mechanics}

In our treatment, spicules are assumed to be axisymmetric, and
individual spicules are 
joined to the central body along the
contour $L$, as illustrated in Fig. 2.
Thus, our calculation involves finding a family of
energy-minimizing axisymmetric spicule shapes and selecting from that
family those shapes consistent with appropriate mechanical boundary
conditions applied along $L$.

Parametrization of the axisymmetric spicule shape is illustrated in
Fig. 2. The variables $z$ and $r$
measure distances along and perpendicular
to the symmetry axis, respectively, while $s$ measures the
arclength from the pole. The function $z(r)$ determines the shape.
$\theta$ is the angle between the local normal and the symmetry
axis; $C_{m}(r)$, $C_{p}(r)$ are the principal curvatures,
\begin{equation}
C_{m}={{d\theta}\over{ds}} \ \  {\mathrm{and}} \ \ C_{p}=
{{\sin\theta}\over r}.
\label{curvatures}
\end{equation}
In calculating the spicule shape, we shall
assume that the relaxed cytoskeleton is locally flat, a good
approximation as long as the number of spicules is large, so that the
spicule size is small compared to $R_{0}$.
Thus, in forming each spicule, a flat circular patch of relaxed
cytoskeleton must be elastically deformed
to fit the spicule contour.  We assume
that this deformation is axisymmetric, so that the center of the
patch remains at the apex of the spicule,
and each point of the patch at relaxed
radius $s_{0}$ maps to a point at arclength $s$ and radius $r$ on
the spicule contour.
The principal extension ratios can then be written,
\begin{equation}
      \lambda_{1}= {r \over s_{0}}, \  \  \
      \lambda_{2}= {ds \over ds_{0}}.
\label{lambdas}
\end{equation}
When these expressions for $\lambda_{1}$ and $\lambda_{2}$
are inserted into Eq.~\ref{elasticenergy} we obtain an explicit
form for the elastic energy in terms of the functions $s_{0}(s)$ and
$r(s)$,
\begin{eqnarray}
F_{\rm el} &=& \pi K_{\alpha} \int s_{0} ds_{0}\left(
{r \over s_{0}}{ds \over ds_{0}} - 1 \right)^{2} \nonumber \\
& & + \pi \mu \int s_{0} ds_{0} \left(
{r \over s_{0}}{ds_{0} \over ds} +
{s_{0} \over r}{ds \over ds_{0}} - 2 \right).
\label{elasticenergy1}
\end{eqnarray}

Using Eqs.~\ref{totalenergy}, \ref{bendingenergy},
\ref{curvatures}, and \ref{elasticenergy1},
we can implement the stationarity
condition for the free-energy functional (\ref{Phi})
with respect to variations of membrane shape and
cytoskeletal strain.  This leads to a set of five
coupled first-order differential equations, which are
written down explicitly in the Appendix. Note that the ADE term in
Eq. \ref{bendingenergy} enters only through the appearance of an
effective spontaneous curvature
\begin{equation}
        C_{0}^{\rm eff}= C_{0} - {\pi \alpha \over D A}
(\Delta A - \Delta A_0),
\label{effectivecurvature}
\end{equation}
which must be determined self-consistently via Eq. \ref{areadifference}.
Although these  Euler-Lagrange equations look complicated, they can be
related to simple mechanical force-balance conditions
in a way that makes their content entirely
transparent (Mukhopadhyay and Wortis, in preparation).
We have used these equations to calculate
(approximate) spicule shapes.

We now turn to the boundary condition where the spicules meet the
sphere (see Fig. 2). Along each element of $L$, the spicule membrane
exerts a tension $\tau_{\parallel}$ in the plane of the membrane and
perpendicular to $L$, and a tension $\tau_{\perp}$ normal to the
plane of the membrane. There is no tension in the third (``shear'')
direction because of the membrane fluidity.  In addition, the membrane
skeleton exerts an independent tension which must be in-plane, since
the skeleton lacks bending rigidity.  In general, these tensions
would be different at different points of $L$; however, in the
axisymmetric approximation, the tensions are uniform around $L$ and
the skeletal tension is directed radially.  To maintain mechanical
equilibrium, these tensions must balance where the spicules meet at
point A.  Note that A is a point of bilateral symmetry between the
two adjacent spicules, so that in-plane tensions, which act in
opposite directions, always balance by
symmetry.  On the other hand, the normal tensions $\tau_{\perp}$
must vanish individually, since they act in the same direction.
The normal tension is related to the isotropic bending moment
per unit length,
\begin{equation}
M = \kappa_{b}(C_{p}+C_{m}- C_{0}^{\rm eff}),
\label{bendingmoment}
\end{equation}
by (Evans and Skalak, 1980; Mukhopadhyay and Wortis, in preparation) 
$\tau_{\perp} = dM/ds$.
Thus, an appropriate boundary condition along $L$ is
\begin{equation}
     {{d(C_{p}+C_{m})}\over{ds}}= 0.
     \label{boundarycondition}
\end{equation}
This is the boundary condition which we shall apply in our calculations.
Of course, this boundary condition is approximate, since the real
spicule is only approximately axisymmetric.  Indeed, with equal logic,
we could argue by symmetry that $\tau_{\perp}$ should vanish at point
B, where three adjacent spicules meet at a radius from the spicule
axis some 15\%  larger than $r_{L}$.
We shall use this observation in Sec. IV to get a rough
measure of the error introduced by the approximation
(\ref{boundarycondition}).

        We now outline the algorithm used to solve numerically the
Euler-Lagrange equations for stationary spiculated shapes.  A
   ``shooting method'' was employed
to integrate the Euler-Lagrange equations from the
apex to the edge of the spicule, the contour $L$ along which
the normal tension $\tau_{\perp}$ vanishes. The set of solutions
can be characterized by five parameters.  Three of them are the global
parameters: the pressure difference $P$, the Lagrange multiplier
$\Sigma$, and the effective spontaneous curvature
$C_{0}^{\rm eff}$, Eq. \ref{effectivecurvature}.
The remaining two
parameters correspond to initial values of variables in the integration
procedure, i.e., the curvature ($C_{m}=C_{p}$)
and the stretching ratios ($\lambda_{1}=\lambda_{2}$) of the
cytoskeleton at the pole.
Integrating the Euler-Lagrange equations to a point
where Eq.~\ref{boundarycondition} is satisfied determines $r_{L}$
and $\theta_{L}$ and, therefore, $R_{0}$ and $n_{s}$
from Eqs.~\ref{geometry} and \ref{spiculenumber}.
The initial conditions may be adjusted iteratively to fit four
parameters, $A$ (Eq.~\ref{area}), $V$ (Eq.~\ref{volume}),
$\overline{\Delta A}_{0}$ (Eq.~\ref{DeltaA0bar}), 
and the cytoskeletal
stretching ratio, $\gamma$, which measures the ratio of the area of the
calculated spicule to that of the corresponding relaxed cytoskeletal
patch (roughly the same as the ratio of the area
of the full echinocyte to that of
the full relaxed membrane skeleton, since the corrections in Eq.
\ref{area} for the spherical segments are small).  We
are left, finally, with
a one-parameter family of mechanical-equilibrium solutions,
which we take to be labelled
by the spicule number $n_{s}$.  The predicted equilibrium
configuration is the
solution $n_{s}^{eq}$ which minimizes the overall energy.
Of course, for the exact solution, $n_{s}$ is a discrete variable:
Probably, there exists, for given initial conditions,
a set of distinct branches of solutions characterized by
different $n_{s}$'s and different arrangements of spicules on the
surface, as discussed more fully in Sec. V.

\subsection{Scaling analysis}

        Before proceeding to solutions, we wish to identify
the important energy and length scales of the problem.
Because we are working at a temperature which is effectively low, one
overall energy scale drops out of the mechanical problem.
We take this
scale to be set by the bending modulus $\kappa_{b}$ ($\kappa_{b}$
and $\bar \kappa$ are similar in magnitude).
There are three length scales:  The first is the overall scale of the
RBC, which we denote $R$.  The second, we may take to be $1/{\bar C}_{0}$
from Eq.~\ref{C0bar} (which is equivalent to using a length based on 
$\overline{\Delta A}_{0}$, Eq.~\ref{DeltaA0bar}); alternatively,
we can take the second length to be $1/C^{\rm eff}_{0}$, where
$C_{0}^{\rm eff}$ is the effective spontaneous curvature from
Eq.~\ref{effectivecurvature}
(we assume that $\bar{C}_{0}$ and $C_{0}^{\rm eff}$ are comparable in 
magnitude). 
The third length is the elastic length scale $\Lambda_{el}$, 
Eq.~\ref{Lambda}.
In general, there may be other length scales associated with the
relaxed cytoskeletal shape; however,
we shall assume that any such lengths are
comparable to $R$.  For the RBC, $R \gg \Lambda_{el}$.  The length
scale $1/{\bar C}_{0}$ is a control parameter whose magnitude can be
tuned across the range defined by $R$ and $\Lambda_{el}$,
resulting in the observed RBC shape classes.

     The size of the RBC sets the total area
and volume in the problem. Note
that the RBC has a relatively smaller volume for a given surface area than
a sphere. If a sphere with the same surface area has a volume
$V_{\rm sphere}$, then the ratio of RBC volume to that of the equivalent
sphere defines a reduced volume,
$v \equiv V_{RBC}/V_{\rm sphere}$.
For the RBC, $v \approx 0.6$, significantly less than unity, which allows
the RBC to assume its sequence of distinct nonspherical shapes.
In its
usual discocytic phase, the red blood cell is
expected to have a magnitude of $1/{\bar C}_{0}$
that is of the order of the red blood cell dimensions
or larger. Tuning ${\bar C}_{0}$ to sufficiently large positive
values generates the sequence of evaginated
echinocytic shapes; tuning it to sufficiently large negative values
generates the sequence of invaginated stomatocytic shapes.

The significance of the elastic length scale $\Lambda_{el}$
is that it determines
how strong the effect of elasticity is
for structures with some characteristic length, say $L$.
To see this, we imagine rescaling the problem
with the energy scale $\kappa _{b}$ and the length scale $L$ in such
a way as to make all the parameters appearing in the Euler-Lagrange
equations dimensionless.  In this way, we arrive at the rescaled
parameters,
\begin{eqnarray}
	\kappa_{b}^{'}&=&1 \nonumber \\
          P^{'}&=&P L^{3}/\kappa_{b} \nonumber \\
	\Sigma^{'}&=&\Sigma L^{2}/\kappa_{b} \nonumber \\
	K_{\alpha}^{'}&=&K_{\alpha} L^{2}/\kappa_{b} \nonumber \\
	\mu^{'}&=&\mu L^{2}/\kappa_{b} \nonumber \\
	C_{0}^{{\rm eff}'}&=&C_{0}^{\rm eff} L.
	\label{rescaling}
\end{eqnarray}
Equivalent rescaling of the variables, $C_{m}^{'}=C_{m} L$, etc.,
leaves the form of the Euler-Lagrange equations unchanged, although
the parameter values (Eq.~\ref{rescaling}) vary with $L$.  For example,
suppose $L$ is the echinocyte spicule radius.
If $L > \Lambda_{el}$, then $\mu^{'}$ is greater
than unity, and the elastic terms from Eq.~\ref{totalenergy}
dominate the bending-energy terms; conversely,
if $L < \Lambda_{el}$,
then $\mu^{'}$ is smaller than one, and the bending-energy terms
dominate.
Thus, generically, elasticity is strong at length
scales larger than $\Lambda_{el}$ and weak at length scales
smaller than $\Lambda_{el}$.
This statement may seem surprising to those who think of the discocyte
as effectively shaped by bending energy.  The key point here is the
plasticity of the membrane skeleton,
albeit on time scales much longer than
the induced shape changes we are discussing:  If the
relaxed membrane skeleton closely matches the
shape preferred by bending energy alone, then (of
course) elastic effects are small.  This is apparently the case for the
resting discocyte (but definitely not for the echinocyte!).

A similar scaling argument gives insight into
the characteristic spicule size in the echinocyte
region.  Suppose we let $L=\Lambda_{el}$ in Eq.~\ref{rescaling}.
At this length scale, the bending and elastic terms 
($\kappa_b^{'}$, $K_{\alpha}^{'}$, and $\mu^{'}$)
are comparable and the only remaining
parameters are $P^{'}$, $\Sigma^{'}$, and ${C}_{0}^{{\rm eff}'}$
(or, equivalently, ${\bar C}^{'}_{0}$). 
If, in addition,
$P^{'}$ and  $\Sigma^{'}$ are small, then the Euler-Lagrange equations
are equivalent to those for the unconstrained minimization of the
combined bending-plus-elastic problem (\ref{totalenergy}).
But, this is a problem which we understand:  When
${C}_{0}^{{\rm eff}'}$ is small in magnitude, 
then the membrane remains flat 
(at the length scale of $\Lambda_{el}$);
on the other hand, when ${C}_{0}^{{\rm eff}'}$ is large in
magnitude,
then bending energy (which dominates the shape at scales below
$\Lambda_{el}$) will favor budded shapes
on the scale of $1/{C}_{0}^{\rm eff}$,
either invaginated or evaginated depending on the sign of
${C}^{\rm eff}_{0}$.  Between these limits, when $\Lambda_{el}$
and $1/{C}^{\rm eff}_{0}$ are comparable, we expect spicules (for
positive ${C}^{\rm eff}_{0}$)
scaled by $1/{C}^{\rm eff}_{0}$, tending
towards smoother shapes
when $\Lambda_{el}$ is somewhat smaller than
$1/{C}^{\rm eff}_{0}$ and towards sharper, more columnar shapes when
$\Lambda_{el}$ is somewhat larger than
$1/{C}^{\rm eff}_{0}$.  Note that, as long as we rescale to the
length $\Lambda_{el}$, the expected spicule shape is predicted to be
independent of cell size $R$  and to depend only on the scaled
value of ${C}^{\rm eff}_{0}$ and the ratio of the two elastic
constants, $\mu$ and $K_{\alpha}$.

The above argument depends on the condition that the scaled
values of $P$ and $\Sigma$ be small at the elastic length scale.
For smooth and flaccid shapes like the discocyte (and when the
cytoskeleton is not significantly stretched or compressed) the
pressure difference $P$ is generated by the bending energy, so
$P^{'}$ is of order unity on the scale of the cell size $R$, and similarly
for $\Sigma^{'}$.  It then follows from Eq.~\ref{rescaling} that
the condition is well satisfied for the discocyte and other nearby
shapes (as we shall see, the initial echinocyte shapes fall into this
class).  Of course, if the system is pushed too hard, then this
condition breaks down and other length scales enter the
spicule-shape problem.

      In summary, the shape of a spicule arises as a result
of the interplay between
the two length scales, $\Lambda_{\rm el}$ and
$1/{C}^{\rm eff}_{0}$.
When $1/{C}^{\rm eff}_{0}$ is positive and much larger than
$\Lambda_{\rm el}$, the effect of elasticity is strong
and we can expect spicules that are at most gentle bumps. It is only
when $1/{C}^{\rm eff}_{0}$ becomes of the order $\Lambda_{el}$ or smaller,
that we may expect sharp spicules. Thus, if we look at the
series of shape transformations, from discocytes
through discocytes with gentle bumps to sharply spiculated
structures, the sharp spicules will first arise with a
radius comparable to
$\Lambda_{el}$. We shall verify this scenario
in the next section by explicit solution for the stable shapes.

\section{RESULTS}

In this section we report echinocyte shapes calculated according to
the program outlined in Sec. III and using the
elastic moduli ($\kappa_{b}$, $\bar\kappa$, $\mu$,
and $K_{\alpha}$) given in
Sec. I and the RBC parameters ($A$ and $V$) given in Sec. II.
The cytoskeletal
stretching ratio $\gamma $ is not known to good accuracy for the RBC.
In this section,
we take $\gamma =1$; however, we have tested values
in the range 0.7--1.2, and
it is one of our results that variation within this range produces no
appreciable changes in the number of spicules or their shape.  The
spontaneous curvature and relaxed area difference are not known and
presumably vary somewhat over a typical RBC population; however, they enter
the mechanics problem only in the combination
$\overline {\Delta a}_{0}$, Eq. \ref{Deltaa0}, which we take as
our principal control parameter.

       Figure 3 displays the equilibrium spicule shapes that
we obtain for a sequence of values of $\overline {\Delta a}_{0}$.
Figures 4 and 5 plot the corresponding calculated number
of spicules $n_{s}$ which appear on the fully developed echinocyte III
at equilibrium
and the corresponding values of
$C_{0}^{\rm eff}$ as functions of $\overline {\Delta a}_0$.

How well do the calculated shapes and
spicule dimensions agree with experimental
observations (Bessis, 1973; Brecher and Bessis, 1972; Chailley et al., 1973).
As shown in Fig. 3, the calculated shapes fall into three distinct
categories.  For $n_{s}$ less than about 20, corresponding to
$\overline {\Delta a}_{0}$ appreciably smaller than
$0.016$, the spicule shapes are broad and
rather hat-shaped, rather different from those seen in experiment.
As $\overline {\Delta a}_{0}$ and $C_{0}^{\rm eff}$ increase,
the number of spicules increases and the individual spicules become smaller,
just as predicted by the scaling argument at the end of Sec.
III.  By the time $\overline {\Delta a}_{0}$ is at or above $0.016$,
for $n_{s}$ in the range 30--60,
the spicule shapes are columnar with
rounded tops, in good general agreement with the kinds of shapes seen in
experiment.  Beyond $n_{s}=70$, the spicules become narrow-necked, a
shape not typically seen in experiment but
illustrating the dominance of the bending energy, as discussed in Sec.
III.

In the ``good,'' central region, the spicule dimensions and
numbers are very comparable to those seen in experiment.  A typical
example ($\overline {\Delta a}_{0}=0.018$) is shown in Fig. 6(a).  For
this case, we find $n_{s}=41$, a radius of the central, spherical
body of $R_{0}=2.57$, and an aspect ratio, height
over width at base (which also corresponds
approximately to the distance between adjacent spicules), of
about 0.8, all of which are in reasonable agreement with
observation.
The only significant discrepancy is that the
spicule height is close to 1.0 $\mu$m, somewhat
larger than the range 0.6--0.8 $\mu$m seen in experiment, as is the
aspect ratio.  We
shall comment on possible reasons for this discrepancy in Sec. V.
We emphasize that these
shapes are a prediction of the model, not an assumption as they are
in other recent work (Igli\v{c}, 1997; Igli\v{c} et al., 1998a, b).
The only other spicule-shape
predictions are those given by Waugh (1996), whose calculation
suggests cross sections that are less steep at the sides and
more pointed at the apex than what is observed.
Whether this is a consequence of the variational form assumed or of the
neglect of stretching elasticity in the model is not clear.
The calculations of Igli\v{c} et al., based on
a postulated spicule shape and assuming skeletal incompressibility
($K_{\alpha}\rightarrow\infty$), produce
somewhat higher
values of the aspect ratio. At a given reduced area difference
$\overline {\Delta a}_{0}$, they find a spicule number which is
larger than ours by almost a factor of two.
We will return to some of these issues in Sec. V.

Before discussing further what we shall argue are the
non-biological regions of hat-like and narrow-necked shapes, we must
review the full echinocyte sequence (Brecher
and Bessis, 1972; Bessis,
1973; Chailley et al., 1973; Mohandas and Feo, 1975).
Experimental observations of RBC's which are
initially smooth, axisymmetric discocytes show that
echinocytosis occurs in three stages.  The first stage (echinocyte I)
corresponds to irregular disc-like shapes with a central dimple.  In
the second stage (echinocyte II) the irregularities evolve into more
or less well defined spicules and the central dimple disappears,
resulting in an oblately elliptical body decorated with spicules.  In
the third stage (echinocyte III) the central body becomes
approximately spherical and the spicules become somewhat
smaller and more numerous.  When pushed beyond the echinocyte III,
some kind of disconnection between the plasma membrane and the
membrane skeleton occurs (see below) and plasma membrane is shed from
the spicule tips into
microscopic exovesicles.  The remaining spherical shape with
pointed protuberances is called a spheroechinocyte (Bessis, 1973).

Our calculation explicitly assumes that the central
body is spherical and is, therefore, appropriate only for
the late stage-III echinocyte.
What we believe happens is that, for $\overline {\Delta a}_{0} \stackrel{\textstyle<}{\sim} 0.016$,
the branch of echinocyte III shapes
becomes energetically unstable to a branch of
echinocyte II shapes. There is no internal way that our calculation
can show that this crossover takes place; however, there are two
pieces of evidence we find convincing.  The first is a simple
estimate:  A separate calculation of the discocyte branch shows that
its shape (and $\Delta A$, for example)
is roughly independent of $\overline {\Delta a}_0$, 
allowing us to estimate its energy as a function
of $\Delta a_0$.  The calculated energy of our echinocyte III branch is
lower than the discocyte branch but increases as 
$\overline {\Delta a}_0$ decreases. 
The crossing point of these two branches at
$\overline {\Delta a}_0 \sim 0.016$ is, thus, an approximate
lower limit to the transition out of the echinocyte III phase.
The actual transition out of the echinocyte III branch
must presumably take place at a somewhat
higher value of $\overline {\Delta a}_0$.  The second piece of
evidence is that, in separate work which we shall report
elsewhere (Lim, Wortis, Boal, and Mukhopadhyay, 2002),
we have done numerical simulations of RBC shapes using the
same set of numerical parameters chosen here, and we find just such a
transition at $\overline {\Delta a}_0
\stackrel{\textstyle<}{\sim} 0.017$ from the echinocyte III
to a branch of echinocyte II shapes.

At the opposite end of the range of ``good'' spicule shapes, our
calculation fails again and, again, for reasons that cannot be
assessed internally.  Here the size of the predicted
spicule--set by the scale of $1/C_{0}^{\rm eff}$--
becomes comparable to the length of the elementary spectrin
tetramer of the membrane skeleton (Liu et al., 1987).
At this length scale, the continuum picture breaks down for the membrane
skeleton (although it remains valid for the lipid component).  What
presumably happens is that a point of instability is reached, where
lipid flow occurs and
the plasma membrane buds outwards in
regions between the cytoskeletal anchors.  Such buds, lacking cytoskeletal
support, are known to be fragile.  They presumably break off,
leading to microvesiculation, membrane loss,
and the observed spheroechinocytosis.  Naturally, this vesiculation
is expected to happen first at the spicule tips, where
the elastic network is maximally expanded and the distance between
anchors is largest.

In nature, the plasma membrane and its cytoskeleton are bound together
by protein anchors; in our model this binding is replaced by the
condition that the bilayer and the elastic network co-exist on the same
mathematical surface.  This requirement means that there is a local
force per unit area or pressure $Q$ with which the skeleton pulls
inward ($Q>0$) or pushes outward ($Q<0$) on the membrane surface (or,
equivalently, with which the membrane pulls outward or pushes inward,
respectively, on the skeleton).
The expression for this pressure is (Mukhopadhyay and Wortis, in preparation)
\begin{eqnarray}
	Q&=& C_{m}\left[K_{\alpha}\left({r \over s}{dr \over ds} - 1\right)
                   + {\mu \over 2}\left(\left(s\over r\right)^{2}
               -\left(ds \over dr\right)^{2}\right)\right] \nonumber \\
       & &+C_{p}\left[K_{\alpha}\left({r \over s}{dr \over ds} - 1\right)
                   - {\mu \over 2}\left(\left(s\over r\right)^{2}
               -\left(ds \over dr\right)^{2}\right)\right]
          	\label{pressure}
\end{eqnarray}
and it is plotted in Fig.~7 as a
function of the radial distance $r$ from the spicule axis
for ${\overline{\Delta a}}_{0}=0.018$. 
As expected, $Q$ is positive
close to the tip of the spicule and negative
around the neck where the skeleton has been compressed.
The magnitude of the
pressure at the spicule tip is of the order of
20 pN/$\mu$m$^{2}$.
It is useful to compare this pressure to the critical disjoining
pressure necessary to break the anchoring and to separate the membrane
skeleton from the bilayer.  This critical pressure has not, to the best of our
knowledge, been measured; nevertheless, it can be estimated.
We take 10 pN as a crude estimate of the force required to extract
quickly a single anchoring protein from the bilayer.
Taking into account the density of anchors, we
estimate a detachment pressure
of about 2000 pN/$\mu$m$^{2}$ (consistent with measurements reported
in Waugh and Bausserman, 1995), two orders of magnitude larger than
what we have calculated above.
This comparison is important, as membrane-cytoskeleton disjoining is
another potential mechanism for spheroechinocytosis, an alternative to
the one discussed in the paragraph above and has, indeed, been proposed
by Igli\v{c} and others (Igli\v{c} et al., 1995, 1996).  We conclude that
such direct disjoining seems unlikely,
unless it is associated with anchoring defects.

We have seen for the human erythrocyte that the fully developed
echinocyte III, with well developed spicules, first appears when
$1/C_{0}^{\rm eff}$ becomes small enough to be comparable to the
elastic length scale $\Lambda_{el}$.  It is a plausible hypothesis
that this is a general connection.  If so, there is a direct relation
between the spicule dimension at the onset of
the echinocyte III and the ratio  (Eq. \ref{Lambda})
of the bending rigidity to the elastic
moduli.  Interestingly, Smith et al. (1982) have studied 
spiculated shapes
for the red cells of a variety of species.  Despite considerable
variation in the volume of the red cells--ranging from the goat RBC, which
is appreciably smaller than the human one, through the elephant seal,
which is appreciably larger--the spicule size is quite stable,
suggesting that the elastic length scale is not strongly variable
from one species to another.  This means, of course, that the number of
spicules on the typical echinocyte is small for the goat ($\sim 10$) and
large for the elephant seal ($\sim 100$), as observed.
To illustrate this point further,
we plot in Fig. 8 spicule shapes at different volumes and
different ratios $K_{\alpha}/\mu$ but with $\Lambda_{el}$ held fixed.
We find that at smaller volume, the spicules sizes do become smaller
but the dependence on volume is weak. The dependence
on the ratio of elastic constants $K_{\alpha}/\mu$ is even weaker.
At a value of $K_{\alpha}$ that is  larger by a factor of 2
($K_\alpha=6 \mu$), the spicules become somewhat broader 
and their number decreases
by around five percent.
Of course, the local stretching of the cytoskeleton both at the
spicule apex and in the neck region close to the base
may be expected to change considerably.

These results are based on the model described in Sec. III, which
assumes spicule axisymmetry, as incorporated in the (approximate)
boundary condition, Eq. \ref{boundarycondition}.  In closing this
section, it may be useful to comment on the reliability of this central
assumption.  We have two
remarks.  The first concerns the boundary condition.  The
neighborhood of a given spicule is clearly not axially symmetric, so
this condition is clearly approximate.  If the calculated shapes
depended sensitively on the precise way that this boundary condition
is applied, then results based upon it would be suspect.  We motivated
the boundary condition ($\tau_{\perp}=0$) by looking at the point
labelled A in Fig. 2, where adjacent spicules are tangent.
With equal logic, we could have argued this boundary
condition  at the point
labelled B.
If we made this choice, how different would the calculated
spicule shapes turn out to be?  We have run some test cases and find
changes in, e.g., the spicule height of less than $1\%$, suggesting
that the results are strongly insensitive to the details of the boundary
condition.  But, this is far from sufficient.  As a second test, we
have recently completed a program 
(Lim et al., in preparation)
of direct simulation of the
energy functional (\ref{totalenergy}) on a triangulated surface.
This program allows us to calculate echinocyte shapes without
assumptions about spicule axisymmetry.  We find echinocyte shapes in
this region of parameter space and can compare the simulated
spicule shapes with those calculated here at the same parameter
values.  Figure 6 shows one such comparison.  Although the base of the
simulated spicule is a little narrower than that given by the
approximate axisymmetric calculation, the overall level of agreement is
excellent.  At these parameter values, the number of spicules in our
calculation is 41, while the number found by simulation is 34.
Similarly, the radius of the central spherical body to which our
spicules are fitted is 2.57 $\mu$m, while that found in the
simulations is about 2.5 $\mu$m (the central body is only 
roughly spherical in this case).  The approximate
calculation presented in this paper appears to be reasonably reliable.

\section{DISCUSSION}

The calculations presented here build on the previous work of Igli\v{c}
et al. (Igli\v{c}, 1997; Igli\v{c} et al., 1998a, b) 
and show that a simple mechanical model of a uniform
composite membrane can
give a good account of observed echinocyte shapes.  The central
control parameter in this process is the effective area difference
(or curvature) of the lipid component, in agreement with the
bilayer-couple hypothesis of Sheetz and Singer (1974).  The role of the
cytoskeletal elasticity is to suppress the formation of narrow-necked
buds which would form in the absence of cytoskeletal shear resistance.
Does this good agreement prove that mechanics (as opposed to
biochemistry) is the sole determinant of echinocyte shape?  Of course,
not; however, it does strongly suggest that simple mechanics will
play an important role in any complete picture.  We end this section
with a brief discussion of some loose ends and future directions.

\subsection{Elastic constants and elastic nonlinearities}

There has been much discussion in the literature of what are the
``correct'' elastic constants to use for the membrane skeleton.  We
have chosen $\mu \approx 2.5 \times 10^{-6}$
J/m$^{2}$ and $K_{\alpha} \approx 3\mu$.  These values are in rough
agreement with recent measurements  at low deformation
(H\'{e}non et al., 1999; Lenormand et al., 2001)
and somewhat lower than those reported in connection with
pipette aspiration of red cells (Evans, 1973a, b; Waugh and Evans, 1979),
where large deformations play an important role.

Discussion of values of the elastic moduli is complicated by the
occurrence of nonlinearities in the elastic response.
At low deformation, $\lambda_{i}=1+\epsilon_{i}$ with
$\epsilon_{i}$ small, there are only two rotational invariants from
which to construct the elastic energy,
${1\over 2}(\epsilon_{1}+\epsilon_{2})^{2}$ and
${1\over 2}(\epsilon_{1}-\epsilon_{2})^{2}$, whose coefficients define
the linear moduli of stretching and shear elasticity.
It is easy to verify that these linear moduli are just $K_{\alpha}$ and
$\mu$, respectively.
However, at large deformation, terms of higher orders in
$(\epsilon_{1}+\epsilon_{2})$ and $(\epsilon_{1}-\epsilon_{2})^{2}$
will generally also appear in the elastic energy, each with its own 
coefficient.
The expression (\ref{elasticenergy}) for the elastic energy makes a
particular choice for each of these coefficients, one which is simply
proportional to the linear moduli.  At small extension ratios, these
higher-order terms are unimportant; however, for the echinocyte shapes,
we find $\epsilon_{i}$ of order unity, so these terms cannot be
neglected.  Unfortunately, present experiments do not constrain the
values of the higher order coefficients significantly, although it is
probably not an accident that the moduli derived from the pipette
aspiration experiments (Evans, 1973a, b), which have extension 
ratios up to nearly 2 (Lee et al., 1999),
are larger than those measured in the linear regime
(Henon et al., 1999; Lenormand et al., 2001).
This suggests a hardening of the elasticity at large deformation,
consistent, for example, with terms in
$(\epsilon_{1}+\epsilon_{2})^{3}$,
$(\epsilon_{1}+\epsilon_{2})(\epsilon_{1}-\epsilon_{2})^{2}$, and so
forth, coming into play.
We will deal with the issue of nonlinearity at more length in a
future publication.  For the present, we simply remark that realistic spicule
formation does require a proper balance between stretching and shear
energies at large deformation.  Our chosen ratio of $K_{\alpha}/\mu =3$
suffices; however, a lower ratio plus some additional nonlinear
hardening would also achieve the same effect and may be more realistic.
Any effect that changes spicule shape significantly
may also be expected to modify
quantitatively the regions of stability of different shape classes.

Previous variational work on echinocyte
shapes by Igli\v{c} and coworkers (Igli\v{c}, 1997; Igli\v{c} et al., 
1998a, b) 
has assumed local incompressibility of the cytoskeleton
($K_{\alpha}\rightarrow\infty$), thus forcing all the elastic energy into the
shear term.  This was done for practical reasons; we are now in a
position to understand the effect of this approximation.  Figure 8(c)
suggests  that increasing $K_{\alpha}$ at fixed $\mu$ has a
modest effect on spicule shape, decreasing the height and the base
radius somewhat without changing the mean radius appreciably.
This is consistent with the generally larger number of spicules
predicted by Igli\v{c} et al. (Igli\v{c}, 1997; Igli\v{c} et al., 1998a, b). 
Because of the limited shapes
available in the variational parametrization, these authors cannot follow
the full spicule-shape evolution exhibited in our Fig. 3.

\subsection{Spicule placement, metastability, and related matters}

Our calculation assumes for the echinocyte regular
spicule placement with six-fold coordination on a
central spherical body.  Of course, this is only an approximation. 
First, there
is the topological requirement for a net undercoordination of 12.
More generally, if the energy minimization problem were to be solved
exactly, we would expect to find several distinct sheets of
locally-minimizing shapes with different spicule numbers $n_{s}$ and
different spicule organization.  For any given set of parameter
values, one such shape would have the lowest energy; however, several
others might have nearby energies. Metastability of a shape would
be expected, since the energy landscape is on the scale of
$\kappa_{b}$, which is large on the scale of $k_{B}T_{room}$. 
This metastability, in turn, would cause hysteresis in the shape 
transformations.

Some  experiments show
that, as the red cell is chemically
cycled back and forth between discocyte and
echinocyte phases, spicules always appear at the same locations on
the cell surface (Furchgott, 1940; Bessis and Prenant, 1972), suggesting
that spicule placement could be intrinsically related to defects
and other inhomogeneities in the cytoskeleton.
This would not happen in a continuum model.
Without wishing to dispute the experimental evidence, we can
only say that our work shows that spicule formation does not require
cytoskeletal inhomogeneity.  Of course, if inhomogeneities exist,
then they will have an effect in selecting the pattern of spicule
placement.

It is not clear a priori whether or not the spicule is
a solitonic object with its own intrinsic scale and, if so, whether
spicules tend to attract or repel at long distance.  Some recent work
(Lim et al., in preparation) has suggested long-range attraction, 
which would mean that,
under appropriate circumstances, spicule clumping might occur, leaving
``bald'' regions on the echinocyte surface, as is seen in some
pictures of echinocyte shapes (see, for example, Igli\v{c} et al., 
1998a).  None of this finer detail is captured in our approximation.

\subsection{Extended shape/phase diagram}

We have focussed on the spontaneous curvature or area difference
$\overline {\Delta a}_{0}$ as the principal parameter
controlling the evolution of
echinocyte shape.  Of course, induced changes in any parameter of the
membrane energy expression Eqs. \ref{totalenergy}--\ref{elasticenergy} 
and the constraints $V$ and $A$, singly or in combination, 
will affect the cell shape. A (bio)chemical modification of the
red-cell environment presumably changes all these parameters to some extent;
it is just that some changes are larger and more important than others.
If there are situations in which laboratory reagents cause important
modification in other parameters--e.g., the elastic constants--then
addition of these reagents will cause
the red cell to be driven along a trajectory in an extended parameter
space.  In such a situation, we would need to consider the
minimum-energy shape (or shapes) as functions of several variables in
an extended shape/phase diagram.  The work surrounding Fig. 8
illustrates the beginning of such a study.

We have summarized in Sec. I the mechanism by which some common reagents
affect the Sheetz-Singer (1974) parameter $\overline {\Delta a}_{0}$.  The
pH effect and the associated glass effect are less readily
related to $\overline {\Delta
a}_{0}$ but may well be primarily related to another dimension of the
generalized phase diagram.  For example, Gedde et al. (1995)
have shown that lipid asymmetry and the presence of
inner leaflet titratable groups, which might be expected to be
associated with $\overline {\Delta a}_{0}$, do not have appreciable
influence on the pH effect; however,
Elgsaeter et al. (1986) and Stokke et al. (1986)
have shown that the membrane skeleton expands in vitro in
response to high cytoplasmic pH.  Now, according to the scaling argument
(Eq.~\ref{cytoscaling}), we know that such expansion is equivalent to
decreasing the shear modulus $\mu$ and, therefore, to increasing the
elastic length scale $\Lambda_{el}$ (\ref{Lambda}).  But,
echinocytosis occurs when $1/C_{0}^{\rm eff}$ shrinks below
$\Lambda_{el}$.  Thus, at fixed $C_{0}^{\rm eff}$ (roughly equivalent
to fixed ${\overline{\Delta a}}_{0}$), expanding the membrane skeleton
could promote echinocytosis, as observed.  It would be premature to claim
that this is the full explanation of the pH effect.  The point is
that, as long as changes in biochemical variables produce homogeneous
changes in the material parameters that appear in Eqs.
\ref{totalenergy}-\ref{elasticenergy}, our model continues to
apply, although the relevant experimental
trajectories may be in a larger parameter space.

\subsection{Membrane composition, inhomogeneity and charge effects}

Description of any effects which lead to inhomogeneity in membrane
properties requires modification of our model.  One such effect
which is certainly present arises from the inhomogeneous distribution of
lipids and/or proteins.  We know that the plasma membrane is
composed of a mixture of lipids and proteins.  The bending modulus $\kappa_{b}$
depends on membrane composition.  As long as the composition remains
homogeneous, the model Eqs.
\ref{totalenergy}--\ref{elasticenergy} holds; but, any mechanism
that produces compositional inhomogeneity on the scale of RBC shape
features would necessitate modification of the model.  Thus, for example, observed raft
formation (Simons and Toomre, 2000; Pralle et al., 2000)
suggests membrane inhomogeneity on the scale of 0.05-0.07 $\mu$m.
When local radii of
curvature become comparable to the inhomogeneity scale, then
inhomogeneity is expected to influence shape.  This may occur for the
echinocyte near the spheroechinocyte limit;
however, there is no evidence that it is a dominant
effect elsewhere.

Other interesting effects involve the coupling of lipid
composition to geometry.  Thus, for example, outer-leaflet lipids 
with large heads
may be expected to segregate preferentially
to spicule tips, as would inner-leaflet lipids
with large tails.  Such an effect could be included in a mechanical
model by adding a composition field (or fields) and including terms
coupling composition to curvature.  Alternatively, one might imagine
mechanisms involving membrane components which preferentially favor (or
avoid) regions around the cytoskeletal anchoring complexes.  These
components would be relatively dilute (concentrated) in regions where
the cytoskeleton was significantly expanded (as it is near the spicule
tip).  To represent this effect, one would need to couple composition
to local cytoskeletal strain.

While these mechanisms may be present to some extent in
the red cell, our calculations suggest they are not in any way
required for spicule formation.

Finally, chemical groups which carry net charge at physiological pH
are common both for lipid headgroups and for cytoskeletal
components.  As long as these charges are compensated by counterions
on scales smaller than the local radii of curvature, they enter the
mechanics only via their effect on the mechanical moduli.  On the
other hand, when relevant lengths become comparable to or smaller than
the Debye length, then charge effects must be treated as nonlocal.

\subsection{Conclusions}

Our calculation provides echinocyte sizes and shapes in excellent 
agreement with
experiment.  Although shapes are calculated under the approximation of
spicule axisymmetry, there is evidence (e.g., Fig. 6) that
this approximation is good.  The approximation does not allow us to
estimate internally the range of echinocyte III stability; however,
extrinsic arguments (Sec. IV) suggest that observable shapes should be
bounded at low $\overline {\Delta a}_0$ in the region where
$1/C_{0}^{\rm eff}$ first becomes smaller than $\Lambda_{el}$ and at
large $\overline {\Delta a}_0$ in the region where the spicule
dimension becomes comparable to the cytoskeletal discreteness.  And,
indeed, outside this region the predicted shapes are no longer seen
experimentally.  While this consistency lends credibility to the
continuum picture, the way in which biochemical influences drive the
control parameters of the model is not addressed.  The Sheetz-Singer
parameter, $\overline {\Delta a}_0$, which measures the area difference
between the lipid monolayers (and the related effect of membrane
spontaneous curvature) does seem to be the dominant control parameter for
some kinds of induced shape change.  On the other hand, other
effects--such as the pH effect--may probe other dimensions of the
shape/phase diagram, e.g., the role of the cytoskeletal
elastic constants and prestress.  We hope that having a mechanical
model with predictive power may eventually help to elucidate
important biochemical questions.  From this perspective, experiments
can now focus on the way in which biochemical probes affect
all the mechanical control parameters, including elastic and cytoskeletal
variables, in addition to $\overline {\Delta a}_0$.

Acknowledgement:

We thank Evan Evans, Ted Steck, and
Narla Mohandas for advice and encouragement, and Dennis Discher
for helpful discussions and suggestions.  Andrew Rutenberg
kindly supplied some important references.  This work was
supported in part by the Natural Sciences and Engineering Research
Council of Canada. RM acknowledges support for a part of the work
from the National Science Foundation under grant number DMR00-96531.  

\section{Appendix}

The Euler-Lagrange equations
obtained by minimizing $\Phi$ may be written down
as the following set of five coupled ordinary
differential equations expressing the variables
$s_{0}$, $r$, $\theta$, $C_{m}$ and $b$ as functions of the arclength
$s$:
\begin{eqnarray}
2 K_{\alpha} r \left( {ds \over ds_{0}} {r \over s_{0}}
-1\right) + \mu s_{0} \left({s_{0} \over r} - {r \over s_{0}}
\left({ds_{0} \over ds}\right)^{2}\right)&+&  2 r \sigma
   \nonumber \\
+ \kappa_{b} r \left[ C_{p}^{2} - 2C_{0}^{\rm eff}C_{p} - C_{m}^{2} \right]
  - b(s) \cos \theta + P r^{2} \sin \theta &=& 0,
\end{eqnarray}
\begin{eqnarray}
{dr \over ds} &=& \cos\theta\\
{d\theta \over ds}&=& C_{m}\\
{{dC_{m}} \over ds}& =& {{\cos \theta \sin\theta}
\over r} - {{C_{m} \cos\theta} \over r}
+{{P r \cos\theta} \over 2\kappa_{b}} + {{b \sin\theta} \over 2r\kappa_{b}}
 \\
{db \over ds}&=& 2 K_{\alpha} \left({ds \over ds_{0}} {r \over s_{0}}
- 1 \right)   \nonumber \\
& & +\mu s_{0}{ds_{0} \over ds} \left({1 \over s_0}
{ds_{0} \over ds} - {s \over r^{2}}{ds \over ds_{0}}\right)
+2 \sigma  \nonumber \\
& &  +\kappa_{b} \left[c_{m}^{2} - 2C_{0}^{\rm eff}C_{m} - C_{p}^{2}\right]
+ 2 p r \sin \theta.
\end{eqnarray}
Here $b(s)$ is a local variable introduced to impose the constraint,
$dr/ds = \cos\theta$, as first suggested by Peterson (1985).
The variables $s$, $s_{0}$, $r$, $\theta$ and
$b$ vanish at the North pole. Note also that the
surface tension $\sigma$ used in these equations
is, in general, shifted from the parameter $\Sigma$
appearing in Eq.~\ref{Phi}.

\pagebreak
\vspace*{0.2in}

\begin{center}
{\bf REFERENCES}
\end{center}
\begin{description}

\item{} Alberts, B., D. Bray, J. Lewis, M. Raff, K. Roberts and J.D. Watson. 1994.
Molecular Biology of the Cell, 3rd ed., Chs. 10 and 16.
Garland Publishing, New York.

\item{} Backman, L. 1986.
Shape control in the human red cell.
{\it J. Cell Sci.} 80:281-298.

\item{} Bennett, V. 1990.
Spectrin-based membrane skeleton: A multipotential adaptor between plasma membrane and cytoplasm.
{\it Physiol. Rev.} 70:1029-1065.

\item{} Bessis, M. 1973.
Living Blood Cells and their Ultrastructure, tr. R.I. Weed.
Springer-Verlag, New York.

\item{} Bessis, M., and M. Prenant. 1972.
Topographie de l'apparition des spicules dans les \'{e}rythrocytes cr\'{e}nel\'{e}s (\'{e}chinocytes).
{\it Nouv. Rev. fr. H\'{e}mat.} 12:351-364.

\item{} Boal, D.H. 1994.
Computer simulation of a model network for the erythrocyte cytoskeleton.
{\it Biophys. J.} 67:521-529.

\item{} Bo\v{z}i\v{c}, B., S. Svetina, B. \v{Z}ek\v{s}, and R.E. Waugh. 1992.
Role of lamellar membrane structure in tether formation from bilayer vesicles.
{\it Biophys. J.} 61:963-973.

\item{} Brecher, G., and M. Bessis. 1972.
Present status of spiculated red cells and their relationship to the discocyte-echinocyte transformation:  A critical review.
{\it Blood} 40:333-344.

\item{} Byers, T.J., and D. Branton. 1985.
Visualization of the protein associations in the erythrocyte membrane skeleton.
{\it Proc. Natl. Acad. Sci. (USA)} 82:6153-6157.

\item{} Canham, P.B. 1970.
The minimum energy of bending as a possible explanation of the biconcave shape of the human red blood cell.
{\it J. Theor. Biol.} 26:61-81.

\item{} Chailley, B., R.I. Weed, P.F. Leblond, and J. Maign\'{e}. 1973.
Formes \'{e}chinocytaires et stomatocytaires du globule rouge.
{\it Nouv. Rev. fr. H\'{e}mat.} 13:71-87.

\item{} Christiansson, A., F.A. Kuypers, B. Roelofsen, J.A.F. Op Den Kamp, and L.L.M. Van Deenen. 1985.
Lipid molecular shape effects erythrocyte morphology: A study involving replacement of native phosphatidylcholine with different
species followed by treatment of cells with sphingomyelinase C or phospholipase A.
{\it J. Cell Biol.} 101:1455-1462.

\item{} Deuling, H.J., and W. Helfrich. 1976.
The curvature elasticity of fluid membranes: A catalogue of vesicle shapes.
{\it J. Phys. (Paris)} 37:1335-1345.

\item{} Deuticke, B. 1968.
Transformation and restoration of biconcave shape of human erythrocytes induced by amphiphilic agents and changes of ionic environment.
{\it Biochim. Biophys. Acta} 163:494-500.

\item{} Discher, D.E., N. Mohandas, and E.A. Evans. 1994.
Molecular maps of red cell deformation: hidden elasticity and in situ 
connectivity.  {\it Science} 266:1032-1035.

\item{} D\"{o}bereiner, H.-G., E. Evans, M. Kraus, U. Seifert, and M. Wortis. 1997.
Mapping vesicle shapes into the phase diagram: A comparison of theory and experiment.
{\it Phys. Rev. E} 55:4458-4474.

\item{} Elgsaeter, A., B.T. Stokke, A. Mikkelsen, and D. Branton. 1986.
The Molecular Basis of Erythrocyte Shape.
{\it Science} 234:1217-1223.

\item{} Evans, E.A. 1973a. A new material concept for the red cell membrane.
{\it Biophys. J.} 13:926-940.

\item{} Evans, E.A. 1973b. New membrane concept applied to the analysis of
fluid shear- and micropipette-deformed red blood cells.
{\it Biophys. J.} 13:941-954.

\item{} Evans, E.A. 1974.
Bending resistance and chemically induced moments in membrane bilayers.
{\it Biophys. J.} 14:923-931.


\item{} Evans, E.A., and R. Skalak. 1980.
Mechanics and Thermodynamics of Biomembranes.
CRC Press, Boca Raton. 

\item{} Ferrell, J.E. Jr., K.J. Lee, and W.H. Huestis. 1985.
Membrane bilayer balance and erythrocyte shape: A quantitative assessment.
{\it Biochemistry} 24:2849-2857.

\item{} Fourcade, B., L. Miao, M. Rao, and M. Wortis. 1994.
Scaling analysis of narrow necks in curvature models of fluid lipid-bilayer vesicles.
{\it Phys. Rev. E} 49:5276-5286.

\item{} Fung, Y.C., and P. Tong. 1968.
Theory of the sphering of red blood cells.
{\it Biophys. J.} 8:175-198.

\item{} Furchgott, R.F. 1940.
Observation on the structure of red cell ghosts.
{\it Cold Spring Harbor Symp. Quant. Biol.} 8:224-232.

\item{} Furchgott, R.F. and E. Ponder. 1940.
Disk-sphere transformation in mammalian red cells II. The nature of the anti-sphering factor.
{\it J. Exp. Biol.} 17:117-127.

\item{} Gedde, M.M., E. Yang, and W.H. Huestis. 1995.
Shape response of human erythrocytes to altered cell pH.
{\it Blood} 4:1595-1599.

\item{} Gedde, M.M., and W.H. Huestis. 1997a.
Membrane potential and human erythrocyte shape.
{\it Biophys. J.} 72:1220-1233.

\item{} Gedde, M.M., D.K. Davis, and W.H. Huestis. 1997b.
Cytoplasmic pH and human erythrocyte shape.
{\it Biophys. J.} 72:1234-1246.

\item{} Gedde, M.M., E. Yang, and W.H. Huestis. 1999.
Resolution of the paradox of red cell shape changes in low and high pH.
{\it Biochim. Biophys. Acta} 1417:246-253.

\item{} Gennis, R.B. 1989.
Biomembranes: Molecular Structure and Function. Chapter 4.
Springer-Verlag, New York.

\item{} Gimsa, J. 1998.
A possible molecular mechanism governing human erythrocyte shape.
{\it Biophys. J.} 75:568-569.

\item{} Gimsa, J., and C. Ried. 1995.
Do band 3 protein conformational changes mediate shape changes of human erythrocytes?
{\it Mol. Membr. Biol.} 12:247-254.

\item{} H\"{a}gerstrand, H., M. Danieluk, M. Bobrowska-H\"{a}gerstrand, 
A. Igli\v{c}, A. Wr\'{o}bel, B. Isomaa, and M. Nikinmaa. 2000.
Influence of band 3 protein absence and skeletal structures on
amphiphile- and Ca$^{2+}$-induced shape alterations in erythrocytes: a
study with lamprey ({\it Lampetra fluviatilis}), trout
({\it Onochorhynchus mykiss}), and human erythrocytes.
{\it Biochim. Biophys. Acta} 1466:125-138.

\item{} Helfrich, W. 1973.
Elastic properties of lipid bilayers: theory and possible experiments.
{\it Z. Naturforsch.} C28:693-703.

\item{} H\'{e}non, S., G. Lenormand, A. Richert, and F. Gallet. 1999.
A new determination of the shear modulus of the human erythrocyte
membrane using optical tweezers.
{\it Biophys. J.} 76:1145-1151.

\item{} Igli\v{c}, A. 1997.
A possible mechanism determining the stability of spiculated red blood cells.
{\it J. Biomech.} 30:35-40.

\item{} Igli\v{c}, A., S. Svetina, and B. \v{Z}ek\v{s}. 1995.
Depletion of membrane skeleton in red blood cell vesicles.
{\it Biophys. J.} 69:274-279.

\item{} Igli\v{c}, A., S. Svetina, and B. \v{Z}ek\v{s}. 1996.
A role of membrane skeleton in discontinuous red blood cell shape transformations.
{\it Cell. Mol. Biol. Lett.} 1:137-144.

\item{} Igli\v{c}, A., V. Kralj-Igli\v{c}, and H. H\"{a}gerstrand. 1998a.
Stability of spiculated red blood cells induced by intercalation of amphiphiles in cell membrane.
{\it Med. Biol. Eng. Comput.} 36:251-255.

\item{} Igli\v{c}, A., V. Kralj-Igli\v{c}, and H. H\"{a}gerstrand. 1998b.
Amphiphile induced echinocyte-spheroechinocyte transformation of red blood cell shape.
{\it Eur. Biophys. J.} 27:335-339.

\item{} Isomaa, B., H. H\"{a}gerstrand, and G. Paatero. 1987.
Shape transformations induced by amphiphiles in erythrocytes.
{\it Biochim. Biophys. Acta} 899:93-103.

\item{} Jay, D.G. 1996.
Role of band 3 in homeostasis and cell shape.
{\it Cell} 86:853-854.


\item{} Landman, K.A. 1984.
A continuum model for a red blood cell transformation: Sphere to crenated sphere.
{\it J. Theor. Biol.} 106:329-351.

\item{} Lange, Y., and J.M. Slayton. 1982.
Interaction of cholesterol and lysophosphatidylcholine in determining red cell shape.
{\it J. Lipid Res.} 23:1121-1127.

\item{} Lange, Y., A. Gough, and T.L. Steck. 1982.
Role of the bilayer in the shape of the isolated erythrocyte membrane.
{\it J. Membr. Biol.} 69:113-123.

\item{} Lee, J.C., D.T. Wong, and D.E. Discher. 1999.  
Direct measures of large, anisotropic strains in deformation of the erythrocyte cytoskeleton. 
{\it Biophys. J.} 77: 853-864.

\item{} Lenormand, G., S. H\'{e}non, A. Richert, J. Sim\'{e}on, and F. Gallet. 2001
Direct measurement of the area expansion and shear moduli of the human red blood cell membrane skeleton.
{\it Biophys. J.} 81:43-56.


\item{} Liu, S.C., L.H. Derick, and J. Palek. 1987.
Visualization of the hexagonal lattice in the erythrocyte membrane skeleton.
{\it J. Cell Biol.} 104:527-536.

\item{} Low, P.S., B.M. Willardson, N. Mohandas, M. Rossi, and S. Shohet. 1991.
Contribution of the band 3-ankyrin interaction to erythrocyte membrane mechanical stability.
{\it Blood} 77:1581-1586.

\item{} Miao, L., U. Seifert, M. Wortis, and H.-G. D\"{o}bereiner. 1994.
Budding transitions of fluid-bilayer vesicles: The effect of area-difference elasticity.
{\it Phys. Rev. E} 49:5389-5407.

\item{} Mohandas, N., and C. Feo. 1975.
A quantitative study of the red cell shape changes produced by anionic and cationic derivatives of phenothiazines.
{\it Blood Cells} 1:375-384.


\item{} Nakao, M., T. Nakao, and S. Yamazoe. 1960.
Adenosine triphosphate and maintenance of shape of the human red cells.
{\it Nature (London)} 187:945-946.

\item{} Nakao, M., T. Nakao, S. Yamazoe, and H. Yoshikawa. 1961.
Adenosine triphosphate and shape of erythrocytes.
{\it J. Biochem.} 49:487-492.

\item{} Nakao, K., T. Wada, T. Kamiyama, M. Nakao, and K. Nagano. 1962.
A direct relationship between adenosine triphosphate-level and {\it in vivo} variability of erythrocytes.
{\it Nature.} 194:877-878.

\item{} Peterson, M.A. 1985.
An instability of the red blood cell shape.
{\it J. Appl. Phys.}  57:1739-1742.

\item{} Ponder, E. 1948.
Hemolysis and Related Phenomena.
Grune and Stratton, New York.

\item{} Ponder, E. 1955.
Red Cell Structure and Its Breakdown.
Springer-Verlag, Vienna.

\item{} Pralle, A., P. Keller, E.L. Florin, K. Simons, and J.K. Horber. 2000.
Sphingolipid-cholesterol rafts diffuse as small entities in the plasma membrane of mammalian cells.
{\it J. Cell Biol.} 148:997-1008.

\item{} Schwarz, S., B. Deuticke, and C.W. Haest. 1999a.
Passive transmembrane redistributions of phospholipids as a determinant of erythrocyte shape change. Studies on electroporated cells.
{\it Mol. Membr. Biol.} 16:247-255.

\item{} Schwarz, S., C.W. Haest, and B. Deuticke. 1999b.
Extensive electroporation abolishes experimentally induced shape transformations of erythrocytes: 
A consequence of phospholipid symmetrization?
{\it Biochim. Biophys. Acta} 1421:361-379.

\item{} Sheetz, M.P. 1979. 
DNase-I-dependent dissociation of erythrocyte cytoskeletons. 
{\it J. Cell Biol.} 81:266-270. 

\item{} Sheetz, M.P., and S.J. Singer. 1974.
Biological membranes as bilayer couples. A molecular mechanism of drug-erythrocyte interactions.
{\it Proc. Natl. Acad. Sci. (USA)} 71:4457-4461.

\item{} Simons, K., and D. Toomre. 2000.
Lipid rafts and signal transduction.
{\it Nat. Rev. Mol. Cell Biol.}  1:31-39.

\item{} Smith, J.E., N. Mohandas, and S.B. Shohet. 1982.
Interaction of amphipathic drugs with erythrocytes from various species.
{\it Am. J. Vet. Res.} 43:1041-1048.

\item{} Steck, T.L. 1989.
{\it In} Cell Shape: Determinants, Regulation, and Regulatory Role.
W.D. Stein, and F. Bronner, editors.
Academic Press, San Diego. 205-246.

\item{} Stokke, B.T., A. Mikkelsen, and A. Elgsaeter. 1986.
Spectrin, human erythrocyte shapes, and mechanochemical properties.
{\it Biophys. J.} 49:319-327.

\item{} Strey, H., M. Peterson, and E. Sackmann. 1995.
Measurement of erythrocyte membrane elasticity by flicker eigenmode 
decomposition. {\it Biophys. J.} 69:478-488. 

\item{} Svetina, S., M. Brumen, and B. \v{Z}ek\v{s}. 1985.
Lipid bilayer elasticity and the bilayer couple interpretation
of red cell shape transformations and lysis.
{\it Stud. Biophys.} 110:177-187.

\item{} Svoboda, K., C.F. Schmidt, D. Branton, and S.M. Block. 1992.
Conformation and elasticity of the isolated red blood cell membrane skeleton.
{\it Biophys. J.} 63:784-793.

\item{} Waugh, R., and E.A. Evans. 1979.
Thermoelasticity of red blood cell membrane.
{\it Biophys. J.} 26:115-131.

\item{} Waugh, R.E., and R.G. Bauserman. 1995.
Physical measurements of bilayer-skeletal separation forces.
{\it Ann. Biomed. Eng.} 23:308-321.


\item{} Waugh, R.E. 1996.
Elastic energy of curvature-driven bump formation on red blood cell membrane.
{\it Biophys. J.} 70:1027-1035.

\item{} Weed, R.I., and B. Chailley. 1972.
Calcium-pH interactions in the production of shape change in erythrocytes.
{\it Nouv. Rev. fr. H\'{e}mat.} 12:775-788.

\item{} Weed, R.I., and B. Chailley. 1973.
{\it In} Red Cell Shape: Physiology, Pathology, Ultrastructure. M. Bessis, R.I. Weed, and P.F. Leblond, editors.
Springer-Verlag, Berlin. 

\item{} Wiese, W., W. Harbich, and W. Helfrich. 1992.
Budding of lipid bilayer vesicles and flat membranes.
{\it J. Phys.: Condens. Matter} 4:1647-1657.

\item{} Wong, P. 1994.
Mechanism of control of erythrocyte shape: A possible relationship to Band 3. 
{\it J. Theor. Biol.} 171:197-205.

\item{} Wong, P. 1999.
A basis of echinocytosis and stomatocytosis in the disc-sphere transformations of the erythrocyte.
{\it J. Theor. Biol.} 196:343-361.

\item{} Wortis, M., M. Jari\'{c}, and U. Seifert. 1997.
Thermal shape fluctuations of fluid-phase phospholipid-bilayer membranes and vesicles.
{\it J. Mol. Liq.} 71:195-207.

\item{} Wortis, M. 1998.
Phospholipid-bilayer vesicle shapes and shape transformations: Theory vs. experiment.
{\it Biol. Skr. Dan. Vidensk. Selsk.} 49:59-63.

\item{} Zarda, P.R., S. Chien, and R. Skalak. 1977.
Elastic deformations of red blood cells.
{\it J. Biomech.} 10:211-221.
 
\end{description}

-------------------------

\newpage
\begin{center}
{\bf FIGURE CAPTIONS}
\end{center}
\begin{description}

\item[Figure 1]
(a)  A budded or vesiculated shape, which would be a low-energy
configuration of a lipid membrane without cytoskeleton at
sufficiently positive spontaneous curvature.  (b)  The spiculated
shape into which (a) is transformed by the elastic-energy
cost of cytoskeletal deformations in the high-shear neck regions.

\item[Figure 2]
Parametrization of a single-spicule shape, showing the boundary $L$
where the spicule joins the spherical central body in constructing
the full echinocyte shape.  The bases of two adjacent spicules are
sketched in to illustrate the points A of spicule tangency and the
point B which is symmetrically situated between three neighboring
spicules.

\item[Figure 3]
Sequence of spicule shapes for increasing values of the reduced effective relaxed area difference
$\overline {\Delta a}_{0}$, Eq. (\ref{Deltaa0}), which measures the combined
effect of spontaneous curvature and additional area in the outer
leaflet in driving the membrane to bend outward.  (a)--(d) correspond
to $\overline {\Delta a}_{0}$ = 0.014, 0.018, 0.020, and 0.022,
respectively.  Note the bar shows the elastic length scale $\Lambda_{el}$.
Note how the spicule sharpens as $\overline {\Delta a}_{0}$
increases and how the neck begins to form when the spicule dimension
(set by $1/C_{0}^{\rm eff}$) falls below $\Lambda_{el}$ so that the
bending energy begins to dominate.

\item[Figure 4]
Calculated equilibrium number of spicules $n_{s}$ plotted as a function of
the reduced effective relaxed area difference $\overline {\Delta a}_{0}$.

\item[Figure 5]
Calculated effective spontaneous curvature
$C_{0}^{\rm eff}$, Eq. (\ref{effectivecurvature}),
plotted as a function of $\overline {\Delta a}_{0}$.

\item[Figure 6]
Calculated spicules at $\overline {\Delta a}_{0}=0.018$.
(a) Shape found in the axisymmetric approximation, as discussed in this paper
($n_{s}=41$, $R_{0}=2.57 \mu$m).  (b)  Shape found by numerical
minimization on a triangulated surface ($n_{s}=34$, $R_{0}=2.5 \mu$m).

\item[Figure 7]
The local pressure $Q$, Eq. (\ref{pressure}), exerted by the cytoskeleton
on the bilayer plotted as a function of radial distance,
for a spicule corresponding to $\overline {\Delta a}_{0}=0.020$.

\item[Figure 8]
Dependence of spicule shape on parameters at $\overline {\Delta a}_{0}=0.018$.
(a) Shape with the standard RBC parameters, as discussed in the text. 
(b) Effect of starting from (a) and increasing $K_{\alpha}$ by a
factor of 2 (i.e. $K_{\alpha}=6 \mu$).  The spicules become slightly broader.
(c) Effect of reducing overall cell dimension
by a factor of two, so that $V \rightarrow V/8$ and $A \rightarrow
A/4$.  This change reduces spicule size by about 10\%.

\end{description}

\newpage
\begin{figure}
\vspace*{1.0truein}
\epsfxsize = 5.in
\epsfysize = 2.in
\centerline{\epsfbox{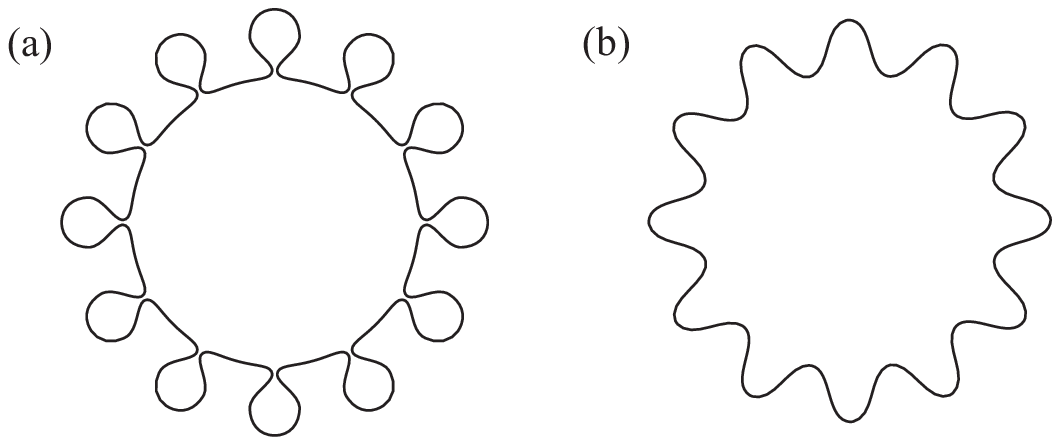}}
\vspace*{1.0truein}
\caption{}
\end{figure}

\begin{figure}
\vspace*{1.0truein}
\epsfxsize = 5.in
\epsfysize = 5.in
\centerline{\epsfbox{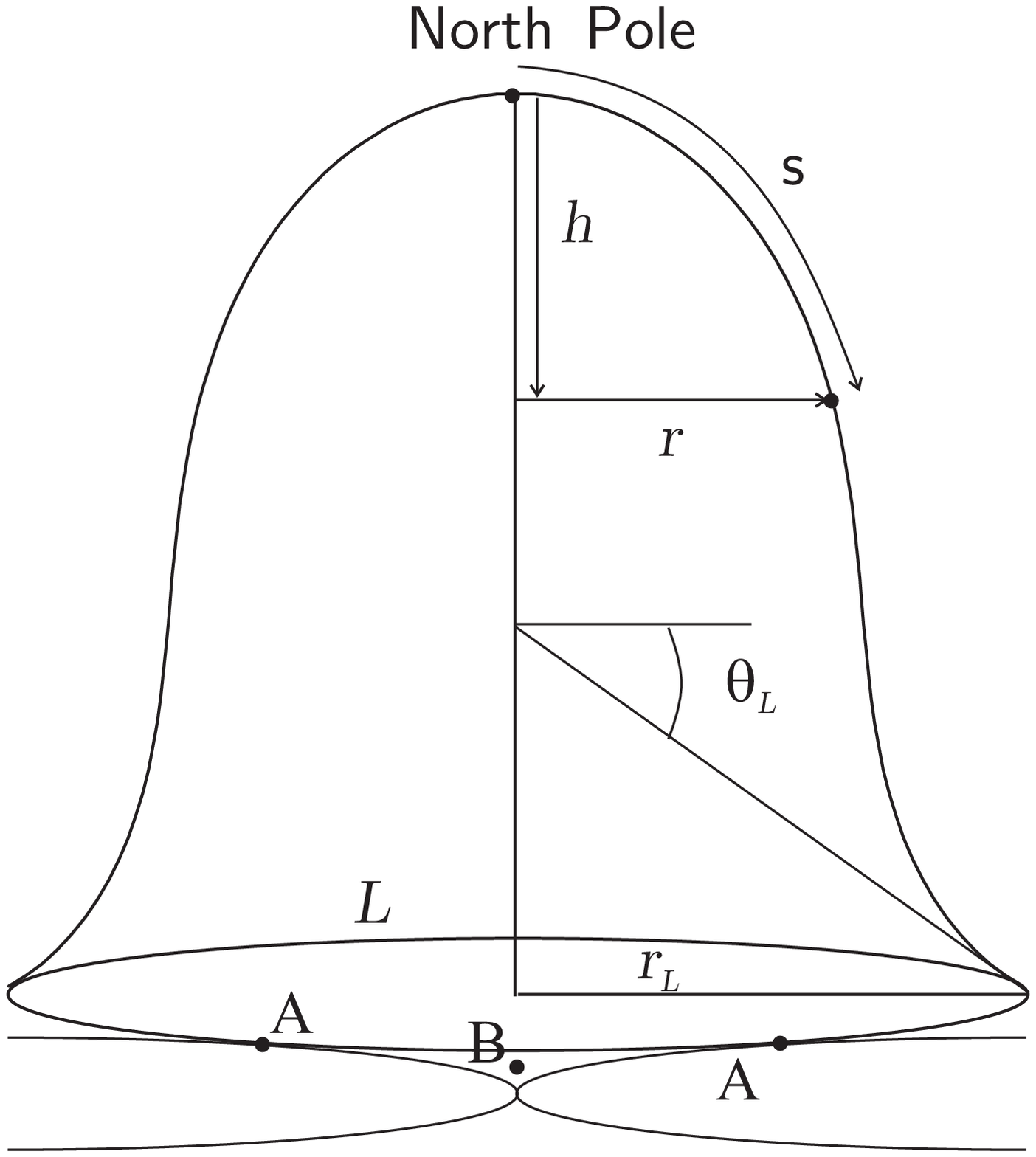}}
\vspace*{1.0truein}
\caption{}
\end{figure}

\newpage
\vspace*{1.0truein}
\begin{figure}
\epsfxsize = 5.in
\epsfysize = 5.in
\centerline{\epsfbox{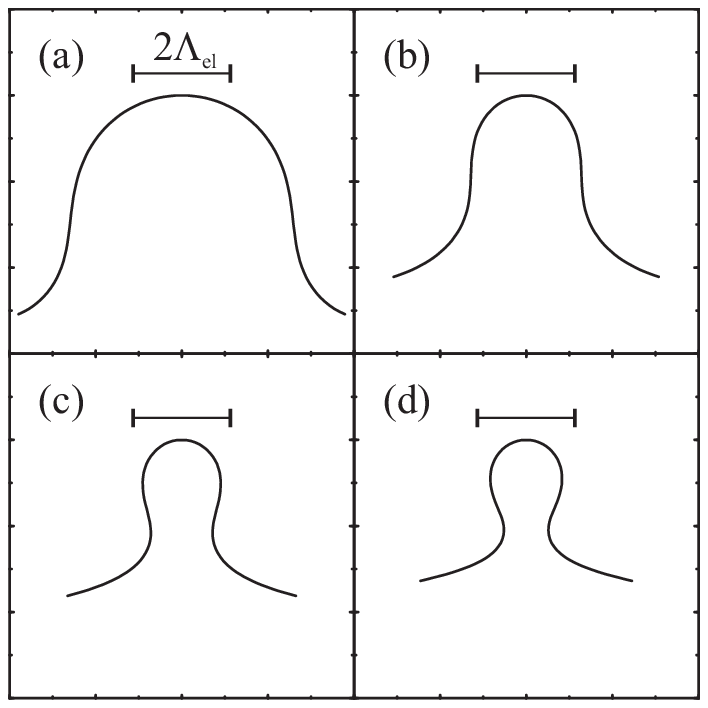}}
\vspace*{1.0truein}
\caption{}
\end{figure}

\newpage
\vspace*{1.0truein}
\begin{figure}
\epsfxsize = 5.in
\epsfysize = 5.in
\centerline{\epsfbox{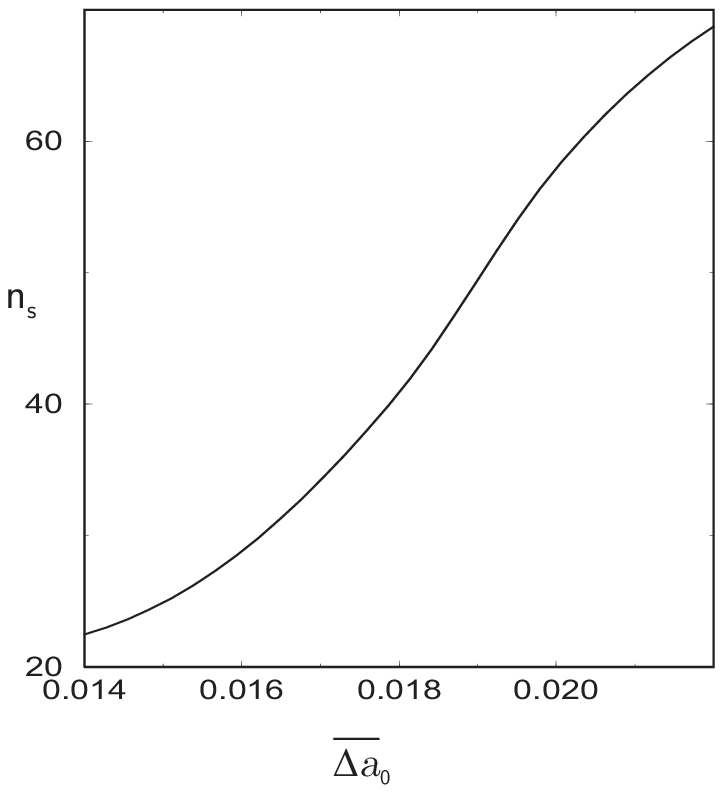}}
\vspace*{1.0truein}
\caption{}
\end{figure}

\newpage
\vspace*{1.0truein}
\begin{figure}
\epsfxsize = 5.in
\epsfysize = 5.in
\centerline{\epsfbox{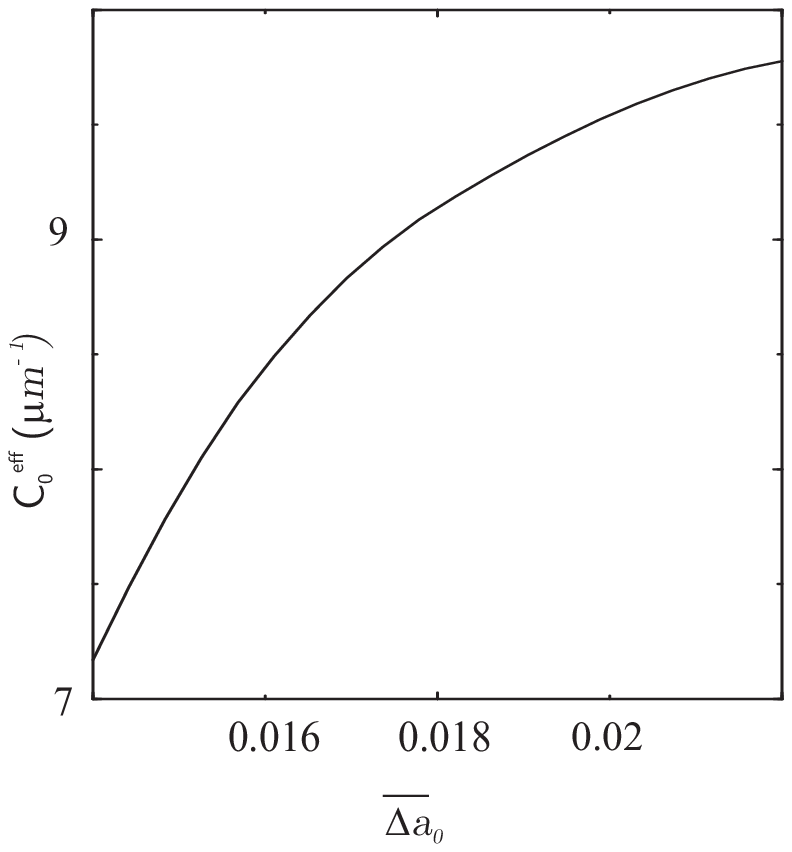}}
\vspace*{1.0truein}
\caption{}
\end{figure}

\newpage
\vspace*{1.0truein}
\begin{figure}
\epsfxsize = 5.in
\epsfysize = 2.5in
\centerline{\epsfbox{cfspicules.eps}}
\vspace*{1.0truein}
\caption{}
\end{figure}


\newpage
\vspace*{1.0truein}
\begin{figure}
\epsfxsize = 5.in
\epsfysize = 5.in
\centerline{\epsfbox{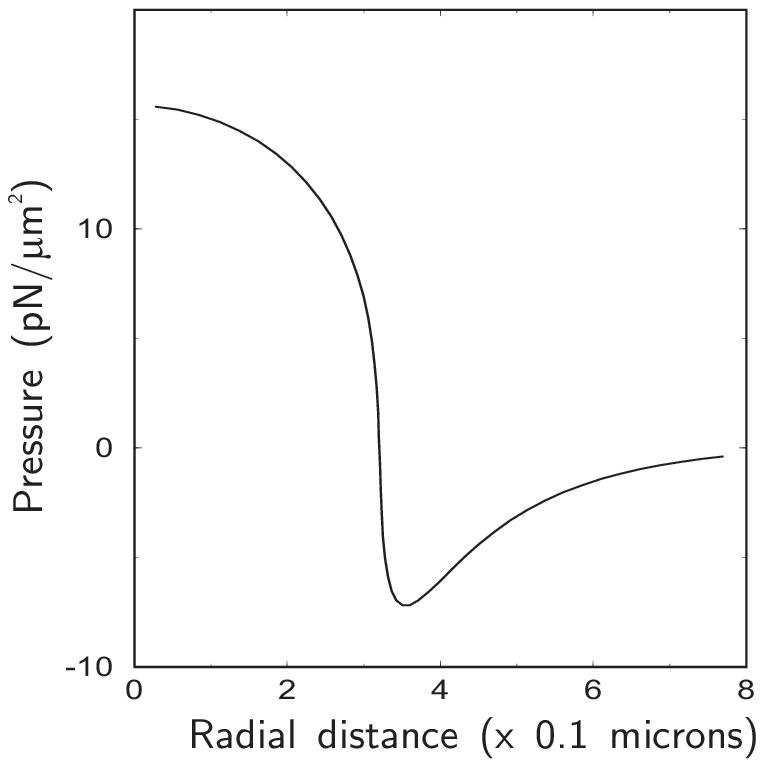}}
\vspace*{1.0truein}
\caption{}
\end{figure}


\newpage
\vspace*{1.0truein}
\begin{figure}
\epsfxsize = 5.in
\epsfysize = 5.in
\centerline{\epsfbox{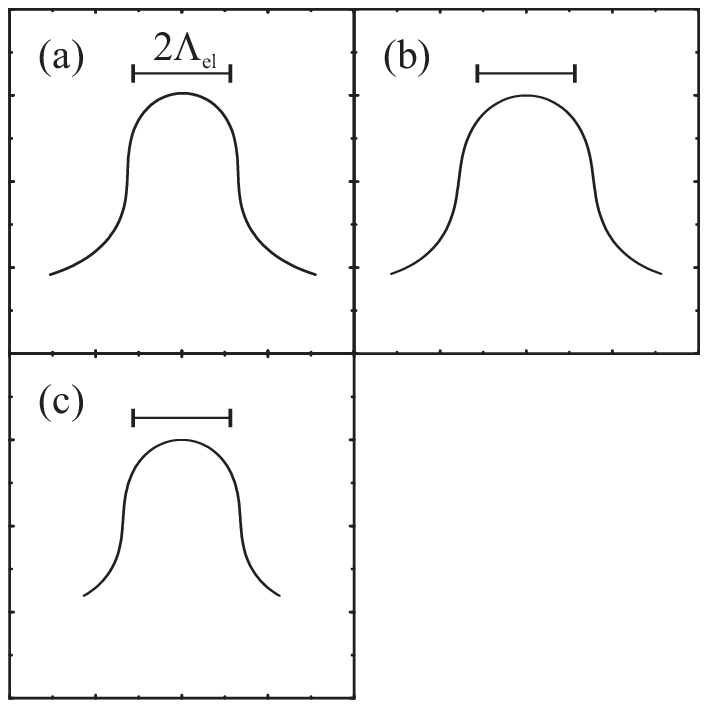}}
\vspace*{1.0truein}
\caption{}
\end{figure}



\end{document}